\documentclass[aip,pof,preprint,showpacs,floatfix]{revtex4-1}
\usepackage{amssymb,amsmath}
\usepackage{natbib}
\usepackage{graphicx}

 \newcommand{\bvec}[1]{\boldsymbol{#1}}
 
 \newcommand{\Cvvec}[1]{\vec{\mathcal{#1}}}
 
 \newcommand{\Mmat}[1]{\mathbb{#1}}

\begin{document}

\title{Hydrodynamic Interactions between Two Forced Objects
of Arbitrary Shape: I Effect on Alignment}

\author{Tomer Goldfriend}
\email{goldfriend@tau.ac.il}
\affiliation{Raymond \& Beverly Sackler School of Physics and Astronomy, Tel Aviv
University, Tel Aviv 69978, Israel}

\author{Haim Diamant} 
\email{hdiamant@tau.ac.il} 
\affiliation{Raymond \& Beverly Sackler School of Chemistry, Tel Aviv
University, Tel Aviv 69978, Israel}

\author{Thomas A.\ Witten} 
\email{t-witten@uchicago.edu}
\affiliation{Department of Physics and James Franck Institute,
University of Chicago, Chicago, Illinois 60637, USA}

\date{\today}

\begin{abstract}
We study the properties and symmetries governing the hydrodynamic
interaction between two identical, arbitrarily shaped objects, driven
through a viscous fluid. We treat analytically the leading (dipolar)
terms of the pair-mobility matrix, affecting the instantaneous
relative linear and angular velocities of the two objects at large separation.
We prove that the instantaneous hydrodynamic interaction linearly degrades 
the alignment of asymmetric objects by an external time-dependent drive 
[Moths and Witten, Phys. Rev. Lett. 110, 028301 (2013)].
The time-dependent effects of hydrodynamic interactions are
explicitly demonstrated through numerically calculated trajectories
of model alignable objects composed of four stokeslets.
In addition to the orientational effect, we find that the
two objects usually repel each other. In this case the mutual degradation
weakens as the two objects move away from each other, and full alignment
is restored at long times.
\end{abstract}


\maketitle

\section{Introduction}
\label{sec:intro}

The dynamics of colloid suspensions is crucially influenced by
flow-mediated correlations~\cite{Happel&Brenner,Russel}. 
While these hydrodynamic interactions (HI)
have an important role
in the dynamics of ambient suspensions at thermal
equilibrium~\cite{Russel}, their effect becomes even more pronounced for
objects driven out of equilibrium,
where the total force acting on each object generates a 
long-ranged flow, decaying as $1/R$ with the distance $R$ between the objects.
A well-known example is colloid
sedimentation, where HI lead to strongly correlated motions and
large-scale dynamic structures~\cite{Ramaswamy2001}.  Various types of
driving, such as electrophoresis,
are widely used to control the transport of colloids and other polyatomic objects~\cite{Russel}.
Theoretical studies of driven colloids traditionally focus on regular particle shapes such as uniform spheres and ellipsoids. The driving
of more asymmetric objects is richer~\cite{Makino&Doi2003,Gonzalez_etal2004,Doi&Makino2005,Makino&Doi2005}
 as it generally includes coupling
between translation and rotation\,---\,when the object is subjected to
a force it also rotates, and when it is under torque it also
translates~\cite{Happel&Brenner}.
The choice of a rotation sense under a unidirectional
force implies a chiral response of the driven object. Such
richer responses can be exploited to obtain {\it``steerable
colloids''}\,---\,objects whose orientation and transport can be
controlled in much more detail. For example, applying a torque by a
rotating uniform magnetic field was used to achieve efficient
transport of chiral magnetic objects~\cite{Morozov&Leshansky2014}. Another
example, which is the main issue of the present work, is the ability
to achieve orientational alignment of asymmetric objects by applying an
external force~\cite{Krapf_etal2009,Moths&Witten2013,Moths&Witten2013b}.

The earlier theoretical works of
Refs.\ \cite{Morozov&Leshansky2014,Krapf_etal2009,Moths&Witten2013,Moths&Witten2013b} dealt with
isolated asymmetric objects in Stokes flow,
which exhibit a chiral response. The object's chiral response 
is encoded in the off-diagonal block of its self-mobility
matrix, referred to as the {\it twist matrix}.
Some objects have a twist matrix that leads them to align one axis in the body with the applied force.
If the twist matrix has only a single real eigenvalue, the object becomes ``axially aligned''
in this way~\cite{Gonzalez_etal2004,Krapf_etal2009}, and the aligning direction is along 
the corresponding eigenvector.
Hence, in the absence of HI and thermal
fluctuations, a set of identical, axially aligning objects reach a
partially aligned state, where all the objects rotate about the same
axis with the same angular velocity, but with an arbitrary phase.
Furthermore, it was shown that, by applying an
appropriate time-dependent forcing, the system can be driven to a
fully aligned state, where all the objects are phase-locked with the force
and rotate in synchrony \cite{Moths&Witten2013,Moths&Witten2013b}.

In view of the above we use throughout this article
the following terminology concerning the response of various objects:
(i) {\it symmetric} objects (such as a uniform sphere); (ii) {\it regular} objects, 
which are asymmetric objects with a vanishing twist matrix (such as a uniform ellipsoid);
(iii) {\it irregular} objects, having a non-vanishing twist matrix; (iv)
{\it axially alignable} objects, which are irregular objects,
whose twist matrix has a single real eigenvalue. 
We note that the twist matrix depends on the position of the forcing point as well.
For example, an ellipsoid whose forcing point is displaced from its centroid, 
i.e., an ellipsoid with a non-uniform mass distribution under gravity, has a non-vanishing 
twist matrix, and generally might be alignable.    

The theoretical groundwork for treating the HI between arbitrary
objects in Stokes flow was laid by Brenner and O'Neill
\cite{BrennerII,Brenner&Oneill1972}.  The theory was subsequently
applied to a pair of particles of various regular shapes
\cite{Goldman_etal1966,Wakiya1965,Felderhof1977,Jeffrey&Onishi1984,Liao&Krueger1980,Kim1985,Kim1986}.
To this one should add  many earlier studies of the collective dynamics of suspensions made of ellipsoids
\cite{Hinch&Leal1972,Brenner1974,Claeys&Brady1993II,Jeffery1922,Davis1991}. 
We note that there are key differences between asymmetric objects, such as ellipsoids,
and the irregular objects studied here. The symmetries of a uniform ellipsoid lead to:
(a) the absence of a translation-rotation coupling for a single object, 
and therefore lack of alignability;
(b) the absence of a $1/R^2$ contribution to the relative velocity 
developed between two such objects at mutual distance $R$.  
Finally, several numerical techniques have been introduced to treat
suspensions of arbitrarily shaped objects
\cite{Karrila_etal1989,Cong&Thien1989,Carrasco&Torre1999,Kutteh2010,Cichocki_etal1994}.

In this work we focus on simple, general properties of the pair HI
between two arbitrarily shaped objects at zero Reynolds number, and
the resulting effect on their orientational alignment. The study of
translational effects will be presented in a separate publication.

The work is made of two distinct parts. The first part treats rigorously 
the instantaneous hydrodynamic interaction, i.e., the pair-mobility matrix.
We use Brenner's analytical framework \cite{BrennerI,BrennerIV}, specializing to
the leading order of the HI in the distance between the objects
(multipole expansion, 
also known as the method of reflections~\cite{Happel&Brenner}).
The second part addresses the time-dependent trajectories of forced objects.
This is a multi-variable, highly non-linear dynamical system exhibiting complex and diverse dynamics.   
In this part we are limited to numerical integration of the objects' trajectories.
We provide typical examples for the time evolution of pairs of stokeslet objects.

We begin by discussing in Sec.\ \ref{sec:general} the general properties
and symmetries of the pair-mobility matrix for two arbitrarily shaped
objects.
In Sec.\ \ref{sec:multipole} we apply a multipole expansion
to the pair-mobility matrix and obtain results for the instantaneous HI
at large distances. 
In Sec.\ \ref{sec:methods} we derive the resulting
properties of stokeslet objects, and in Sec.\ \ref{sec:alignment} we
use them to perform numerical time integration for the evolution of
object pairs and their alignment. Finally, in Sec.\ \ref{sec:discuss}, we
discuss several consequences of our results.

\section{Pair-Mobility Matrix: General Considerations} 
\label{sec:general}

\subsection{Structure of the Pair-Mobility Matrix}
\label{sec:structure}

The kinematics of a rigid object is represented by a translational
velocity $\vec{V}$, which refers to an arbitrary reference point
rigidly affixed to the object, and an angular velocity
$\vec{\omega}$.  We designate the reference point as \emph{the origin}
of the object.  Note that the angular velocity of the object is
independent of the choice of its origin, and that the origin does not
necessarily lie on the instantaneous axis of rotation of the object.

Consider two arbitrarily shaped rigid objects, $a$ and $b$, with typical
size $l$, subject to external forces and torques $\vec{F}^a$,
$\vec{F}^b$ and $\vec{\tau}^a$, $\vec{\tau}^b$ in an unbounded,
otherwise quiescent fluid of viscosity $\eta$. In the creeping flow
regime, the objects respond with linear and angular velocities to the
external forces and torques through a $12 \times 12$
\emph{pair-mobility matrix}, 
\begin{equation}
   \begin{pmatrix}
  \Cvvec{V}^a \\
  \Cvvec{V}^b 
 \end{pmatrix} =
\frac{1}{\eta l}
 \begin{pmatrix}
  \Mmat{M}^{aa}	& \Mmat{M}^{ab}\\
  \Mmat{M}^{ba} & \Mmat{M}^{bb}
 \end{pmatrix}
   \begin{pmatrix}
  \Cvvec{F}^a \\
  \Cvvec{F}^b 
 \end{pmatrix},
\end{equation}
where we define \emph{generalized velocity} and \emph{generalized
  force} 6-vectors, $\Cvvec{V}^x=( \vec{V}^x, l \vec{\omega}^x )^T$
and $\Cvvec{F}^x= (\vec{F}^x, \vec{\tau^x}/l )^T$ for $x=a,b$. The
diagonal blocks, $\Mmat{M}^{aa}$ and $\Mmat{M}^{bb}$, correspond to
the self-mobilities of the objects (which nevertheless depend
on the configuration of both objects). The off-diagonal blocks,
$\Mmat{M}^{ab}$ and $\Mmat{M}^{ba}$, describe the pair hydrodynamic
interaction. We hereafter omit the factor $(\eta l)^{-1}$ (i.e., set $\eta l=1$).
This, together with the representation of the generalized forces and velocities, make
$\Mmat{M}$ dimensionless and dependent on the geometry alone.
Throughout the text we designate 6-vectors and matrices
with calligraphic font and blackboard-bold letters, respectively.
A detailed description of the notation used in the article is given in Appendix~\ref{sec:notation}.   

Since $\vec{V}$ and $\vec{\tau}$ depend on the choice of object origins, 
so does the pair-mobility matrix. The transformation between pair-mobility
matrices corresponding to different origins is given in Appendix~\ref{sec:origin}.

The pair-mobility matrix is a function of the objects' geometries,
their orientations, and the vector connecting their origins, indicated
hereafter by $\vec{R}$. (We define the direction of $\vec{R}$ from the
origin of object $b$ to the origin of object $a$.) The geometry of
object $x$ is denoted by $\bvec{r}^{x}$. For example, if the object
consists of a discrete set of $N_x$ stokeslets (see Sec.~\ref{sec:stokeslet_properties}),
then $\bvec{r}^{x}$ is a $3N_x$-vector specifying the positions of the
stokeslets; otherwise, it represents the surface of the object. 

The pair-mobility matrix is positive-definite and
symmetric~\cite{Condiff&Dahler1966,Happel&Brenner,Landau&Lifshitz}. Hence,
$\Mmat{M}^{ab}=(\Mmat{M}^{ba})^T$, and the self-blocks can be written
as
$$
  \Mmat{M}^{xx} = \begin{pmatrix}
  \Mmat{A}^{xx} & (\Mmat{T}^{xx})^T \\
  \Mmat{T}^{xx} & \Mmat{S}^{xx}
  \end{pmatrix} .
$$ 
As in the analysis for isolated objects \cite{Krapf_etal2009}, the
self-mobility matrix contains the following $3\times 3$ blocks: the
alacrity matrix $\Mmat{A}$ (translational response to force); the
screw matrix $\Mmat{S}$ (rotational response to torque); and the twist
matrix $\Mmat{T}$ (translation--rotation coupling). The twist matrix
characterizes the chiral response of the object (the sense of rotation
under a force). In the present article we deal with alignable objects,
whose individual $\Mmat{T}$ is necessarily non-vanishing. Furthermore,
in the case of a pair of objects, the presence of the other object
makes the self-twist matrix, $\Mmat{T}^{xx}$, differ from the
single-object one. As to the off-diagonal blocks of the pair-mobility
matrix, the symmetry of $\Mmat{M}$ implies the following structure:
$$
  \Mmat{M}^{ab} = \begin{pmatrix}
  \Mmat{A}^{ab} & (\Mmat{T}^{ba})^T \\
  \Mmat{T}^{ab} & \Mmat{S}^{ab}
  \end{pmatrix},\ \ \ 
  \Mmat{M}^{ba} = \begin{pmatrix}
  (\Mmat{A}^{ab})^T & (\Mmat{T}^{ab})^T \\
  \Mmat{T}^{ba} & (\Mmat{S}^{ab})^T
  \end{pmatrix} .
$$

\subsection{Further Symmetries of the Pair-Mobility Matrix}
\label{sec:symmetry}

The discussion in the preceding subsection has been for a general pair of objects,
which are not necessarily identical.
In the present subsection, we focus on the case in which the two objects are
\emph{identical in shape and orientation}, i.e.,
$\bvec{r}^a=\bvec{r}^b\equiv\bvec{r}$.  Our goal is to understand what
the instantaneous relative velocities (linear and angular) between the
two objects are, when the objects are subjected to the same external
forcing. The restriction to identical objects makes $\Mmat{M}$ invariant under
exchange of objects. This additional symmetry is made of two
operations: interchanging the blocks
$\Mmat{M}^{aa}\leftrightarrow\Mmat{M}^{bb}$ and
$\Mmat{M}^{ab}\leftrightarrow\Mmat{M}^{ba}$; and inversion of $\vec{R}$.
That is,
\begin{equation}
 \Mmat{M} (\bvec{r},\vec{R})=
 \Mmat{E}  \Mmat{M} (\bvec{r},-\vec{R})  \Mmat{E}^{-1} ,
\label{eq:switch}
\end{equation}
where $\Mmat{E}$ is a $12\times12$ 
matrix which interchanges the objects,
$$
\Mmat{E}=
 \begin{pmatrix}
  0	& \Mmat{I}_{6\times6}\\
  \Mmat{I}_{6\times6} & 0
 \end{pmatrix},
$$ 
with $\Mmat{I}_{6\times6}$ denoting the $6\times 6$ identity matrix. 

The symmetry to object exchange, when combined with the
parity of $\Mmat{M}$ 
(i.e., whether it remains the same or changes sign)
 under $\vec{R}$-inversion,  
\footnote{\setcounter{footnote}{1} Parity does not mean here symmetry under full spatial inversion,
as such an operation would turn the chiral objects
into their enantiomers; rather, we mean here symmetry under the inversion of $\vec{R}$}
has important consequences for the effect of hydrodynamic interactions on
alignment. If $\Mmat{M}$ has a definite parity one can determine what
the relative response of the objects to forcing is\,---\,i.e., whether
they attain the same or the opposite linear and angular velocities. If
the term is symmetric to inversion, the velocities would be identical,
and if it is antisymmetric, they would be opposite. This is because
\begin{equation}
 \begin{pmatrix}
  \Mmat{M}^{aa}(\vec{R})	& \Mmat{M}^{ab}(\vec{R})\\
  \Mmat{M}^{ba}(\vec{R}) & \Mmat{M}^{bb}(\vec{R})
 \end{pmatrix}=
\pm 
 \begin{pmatrix}
  \Mmat{M}^{aa}(-\vec{R})	& \Mmat{M}^{ab}(-\vec{R})\\
  \Mmat{M}^{ba}(-\vec{R}) & \Mmat{M}^{bb}(-\vec{R})
 \end{pmatrix}
=
\pm
 \begin{pmatrix}
  \Mmat{M}^{bb}(\vec{R})	& \Mmat{M}^{ba}(\vec{R})\\
  \Mmat{M}^{ab}(\vec{R}) & \Mmat{M}^{aa}(\vec{R})
 \end{pmatrix},
\label{eq:even_odd}
\end{equation}
where the second equality comes from the response to exchange of
objects, Eq.~\eqref{eq:switch}. Consequently, under identical forcing
of the two objects one finds,
\begin{equation}
 \Cvvec{V}^a=\left(\Mmat{M}^{aa}+\Mmat{M}^{ab} \right)\Cvvec{F}
= \pm \left(\Mmat{M}^{bb}+\Mmat{M}^{ba} \right)\Cvvec{F}=\pm \Cvvec{V}^b.
\end{equation}
Thus, since any $\Mmat{M}$ can be decomposed into even and odd terms,
we find that only the odd ones cause relative motions of the two
objects.

The pair-mobility as a whole, however, never has a definite parity
under $\vec{R}$-inversion, i.e., it is made of both even and odd
terms. This becomes clear when $\Mmat{M}(\bvec{r},\vec{R})$ is
expanded in small $l/R$, i.e., in multipoles. A general discussion of
the parity of each multipole term is given in the next section. For now, let
us consider those two leading multipoles which are independent of the
objects' shape, and therefore always exist. The monopole--monopole
interaction (Oseen tensor), which is the leading term in
$\Mmat{A}^{ab}$ making particle $a$ translate due to the force on
particle $b$, is symmetric under $\vec{R}$-inversion. The part of the
monopole--dipole interaction causing the second object to rotate due
to the force on the first, i.e., the leading term in $\Mmat{T}^{ab}$,
is antisymmetric. For example, even the most symmetric pair of
objects\,---\,two spheres\,---\,has an $\vec{R}$-symmetric
$\Mmat{A}^{ab}$, leading to zero relative velocity, and an
$\vec{R}$-antisymmetric $\Mmat{T}^{ab}$, causing them to rotate with
opposite senses \cite{Happel&Brenner}. Thus, for a general object, the highest
order which maintains $\Mmat{M}$ of definite parity is the monopole
$1/R$ Oseen one, which is even. (The self-blocks are constant up to
order $1/R^4$; see below.)

From this discussion we can immediately conclude that, to leading
order in the separation of two identical, fully aligned objects, {\em their 
instantaneous hydrodynamic interaction
  must linearly degrade the alignment}. The leading degrading term comes from
$\Mmat{T}^{ab}$, their rotational response to force, and is of order $1/R^2$.
It is worthwhile to note again that such a rotational response is present as well for a pair
of uniform spheres or ellipsoids; yet, such regular objects are not alignable to begin with.

The relation between object-exchange symmetry and the symmetry of the
linear-velocity response is intimately related to the issue of
hydrodynamic pseudo-potentials \cite{Squires2001}, which will be discussed
in detail in a forthcoming publication.

\section{Far-Field Interaction: Multipole Expansion} 
\label{sec:multipole} 

There are two characteristic length scales in our problem: the typical
size of the objects, $l$, and the distance between them,
$R=|\vec{R}|$. If $l \ll R$, we can write the pair-mobility matrix as
a power series in $(l/R)$,
$$
\Mmat{M}=\Mmat{M}_{(0)}+\Mmat{M}_{(1)}+\Mmat{M}_{(2)} + \dots,
$$ 
where $\Mmat{M}_{(n)} \sim  (l/R)^n$. The analysis of this expansion as given below
holds for any pair of objects, whether identical or not.
The zeroth order,
$\Mmat{M}_{(0)}$, is a block diagonal matrix which is made of the self-mobilities of the two 
non-interacting objects. (These should be distinguished from $\Mmat{M}^{aa}$ and
$\Mmat{M}^{bb}$, the self-mobilities of the interacting objects.)

The hydrodynamic multipole expansion (also known as the method of reflections)
is based on the Green's function
of Stokes flow, the Oseen tensor \cite{Happel&Brenner},
given in our units ($\eta l=1$) by
\begin{equation}
 \Mmat{G}_{ij} (\vec{r})=
\frac{1}{8\pi} \frac{l}{r} 
\left( \delta_{ij} + \frac{r_i r_j}{r^2} \right),
\label{eq:Oseen}
\end{equation}
which is a symmetric $3\times 3$ tensor, invariant under
$\vec{r}$-inversion. A point force at $\vec{r}_0$,
$\delta(\vec{r}-\vec{r}_0) \vec{f}$, generates a velocity field
$\vec{u}(\vec{r})=\Mmat{G}(\vec{r}-\vec{r}_0)\cdot \vec{f}$.

We obtain two general results concerning the multipoles of the
hydrodynamic interaction between two arbitrary objects. The two
objects need not be identical. The proofs are given in
Appendix~\ref{sec:proofs}. \footnote{In fact, these results are not
  special to the hydrodynamic interaction but can be similarly proven
  for any multipole expansion. As such, they were most probably
  derived before.}
\begin{enumerate}
{\it
\item The leading interaction multipole in the self-blocks of the
pair-mobility matrix is $n=4$.  
That is, any response of one object to forces on itself, owing to the other object,
must fall off with distance R between the objects at least as fast as $R^{-4}$.
\item The $n$th multipole has self-blocks of $(-1)^n$ parity, and coupling blocks
of the opposite, $(-1)^{n+1}$ parity.
Thus, e.g., the leading term in $\Mmat{M}^{aa}$, proportional to $R^{-4}$, is invariant under $\vec{R}$-inversion,
and the $R^{-4}$ part of $\Mmat{M}^{ab}$ changes sign under $\vec{R}$-inversion.  
Likewise for the multipole varying as $\vec{R}^{-5}$,
the $\Mmat{M}^{aa}$ changes sign under $\vec{R}$-inversion while $\Mmat{M}^{ab}$ remains invariant.\cite{Note1}
}
\end{enumerate}
These statements pertain to the mobility matrix. As to the propulsion
matrix (the inverse of the mobility matrix), the leading correction to the
self-block becomes $\sim 1/R^2$, and the second statement concerning parity remains intact.

We now consider for a moment two identical objects and specialize to the
first and second multipoles, i.e., the hydrodynamic interaction up to
order $1/R^2$. The discussion in the preceding and current sections
implies the following form of the two leading terms in the
pair-mobility matrix:
\begin{equation}
\Mmat{M}_{(1)} =
	\begin{pmatrix}
		  0	& \Mmat{M}_{(1)}^{ab}\\
		  \Mmat{M}_{(1)}^{ab} & 0
	\end{pmatrix},\qquad
 \Mmat{M}_{(2)}=
\begin{pmatrix}
  0 &  \Mmat{M}_{(2)}^{ab}\\
  -\Mmat{M}_{(2)}^{ab} & 0\\
\end{pmatrix}.
\label{eq:M1_M2}
\end{equation}
In more detail: there are no first- and second-order corrections to
the objects' self-mobility. Hence, these two multipoles have definite
parities\,---\,the first is even, and the second is odd.
Consequently, the first multipole does not cause any relative motion
of the two objects, whereas the second mutipole makes them translate
and rotate in opposite linear and angular velocities.

\begin{figure}
\centerline{\resizebox{0.8\textwidth}{!}{\includegraphics[viewport= 0 -1 551 360]{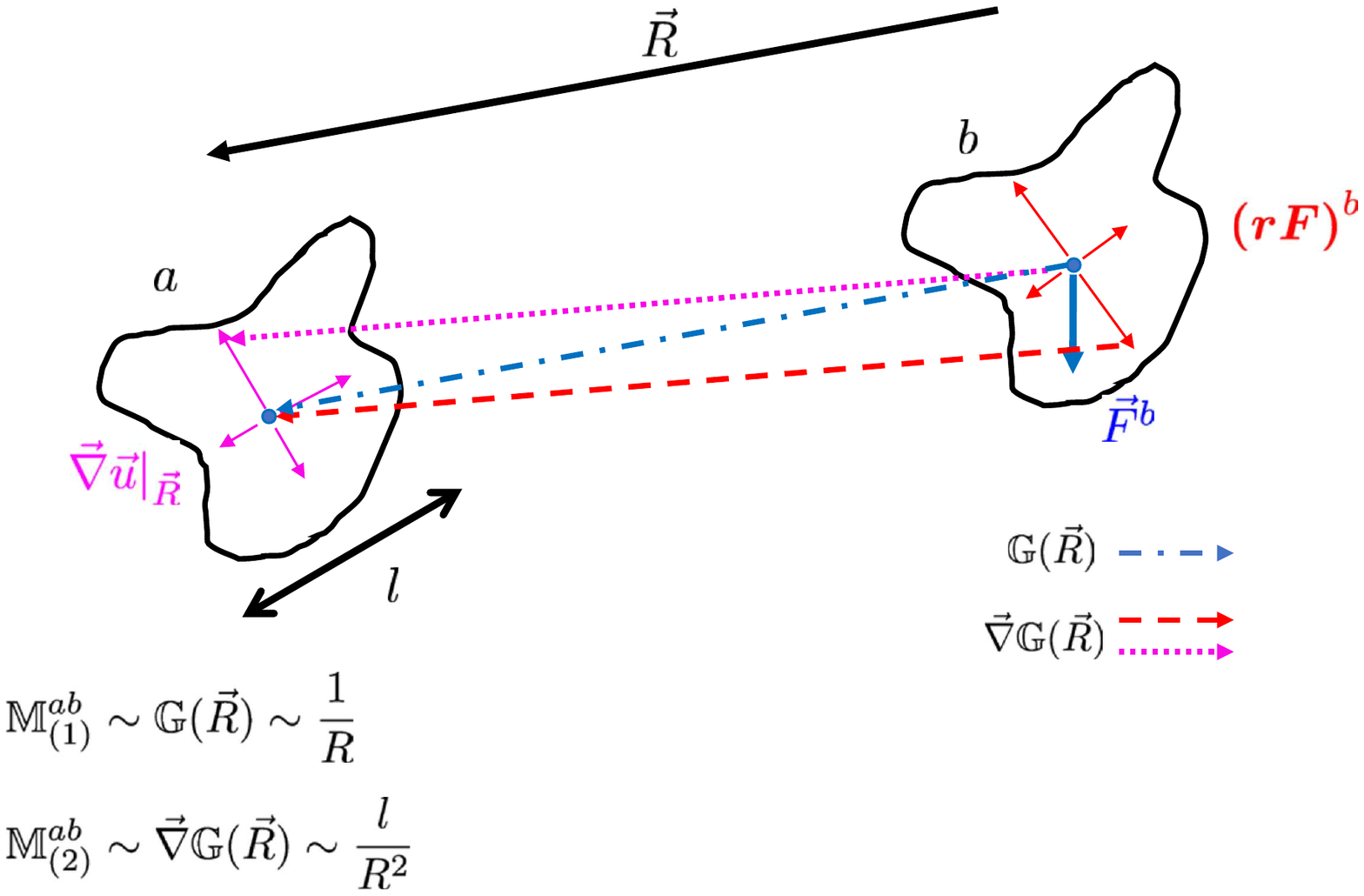}}}
\caption[]{ 
Illustration of the two leading orders of the hydrodynamic interaction between two forced objects.
The leading term in the pair-mobility matrix (light blue/dash-dotted arrow between
the objects' origins), decaying as $1/R$,
comes from the point-like response of object $a$ to the local flow caused by the force monopole on object $b$
(blue/thick arrow). 
The next-order term, decaying as $l/R^2$, has two contributions:
(i) The point-like response of object $a$ to the local flow caused by the force dipole on object $b$
(red/dashed arrow from the red/thin arrows at object $b$ to the origin of $a$).
(ii) The response of object $a$ to the local flow gradient caused by the force monopole on object $b$
(magenta/dotted arrow from the origin of $b$ to the magenta/thin arrows at object $a$).}
\label{fig:multipole}
\end{figure}

The essential characteristics of the first two multipoles
are schematically illustrated in Fig.~\ref{fig:multipole}.
The first multipole arises directly from the Green's function,
\begin{equation}
\Mmat{M}^{ab}_{(1)}=
\begin{pmatrix}
  \Mmat{G}(\vec{R}) & 0\\
  0 & 0
\end{pmatrix},
\label{eq:M1}
\end{equation}
where $\Mmat{G}(\vec{R})$ is the Oseen tensor, given in
Eq.~(\ref{eq:Oseen}).

In the interaction described by the second multipole one object sees
the other as a point,  see Fig.~\ref{fig:multipole}. Accordingly, this term contains two types of
interaction: (1) the response of object $a$ to the non-uniformity of
the flow due to the force monopole at object $b$ (regarded as a
point); (2) the advection of object $a$ (regarded as a point) by the
flow due to the force dipole acting at object $b$. These two effects
are both proportional to $\vec{\nabla}\Mmat{G}(\vec{R})\sim
1/R^2$. Each can be written as a product of a tensor which arises from
the medium alone, through derivatives of the Oseen tensor
$\vec{\nabla}\Mmat{G}(\vec{R})$, and another tensor which depends on
the objects' geometry. The second-order correction to the velocity of
object $a$ is given by the sum of these two effects,
each expressed in terms of a coupling tensor $\Theta$ and an object tensor $\Phi$
\begin{eqnarray}
  \Cvvec{V}^a_{(2)} &=& \Mmat{M}^{ab}_{(2)} \cdot \Cvvec{F}^b \nonumber\\
  \Mmat{M}^{ab}_{(2)} &=& \Phi^a : \Theta(\vec{R}) 
  - \Theta^T(\vec{R}) : \tilde{\Phi}^b,
\label{eq:V2a}
\end{eqnarray}
where the double dot notation denotes a contraction over two indices.
Equation~(\ref{eq:V2a}) contains three tensors of rank 3, denoted by capital Greek letters. 
The first,
$\Phi$,  with dimensions $6\times 3\times 3$, gives the generalized velocity of the object in linear
response to the velocity gradient of the flow in which it is
embedded. The second, $\tilde{\Phi}$, having dimensions $3\times 3\times 6$, gives the force dipole acting on
the fluid around the object's origin in linear response to the
generalized force acting on it.  Both $\Phi$ and $\tilde{\Phi}$ depend
on the objects' geometry alone~\footnote{These tensors are related to
  the two introduced by Brenner~\cite{BrennerIV}. Brenner's tensors
  give the force and torque exerted on an object in linear response to
  a flow gradient in which it is embedded. Our $\Phi$ is related to
  these two via the individual self-mobility matrix.}. The third tensor, $\Theta$,
	with dimensions $3\times 3\times 6$,
describes the coupling of these object responses through the fluid.
It is given by
\begin{equation}
	\Theta_{skj}(\vec{R})  \equiv 
	  \left\{
  	\begin{array}{ll}
        \partial_s \Mmat{G}_{kj}(\vec{r})|_{\vec{R}}\ \ & j=1,2,3 \\
				0\ \ & j=4,5,6.	
        \end{array} \right.
\end{equation}
Repeating the same procedure for
$\Cvvec{V}^b$ in response to $\Cvvec{F}^a$ while using the odd parity
of $\Theta$, we get
\begin{equation}
  \Mmat{M}^{ba}_{(2)} = \Theta^T(\vec{R}) : \tilde{\Phi}^a 
  - \Phi^b : \Theta(\vec{R}).
\label{eq:V2b}
\end{equation}

The tensors $\Phi$ and $\tilde{\Phi}$ are not independent~\cite{Kim&Karrila}. We now show
that $\Phi=\tilde{\Phi}^T$. The symmetry of $\Mmat{M}$ implies that
each multipole is also a symmetric matrix. Using Eqs.~(\ref{eq:V2a})
and (\ref{eq:V2b}) and equating $(\Mmat{M}^{ba}_{(2)})^T =
\Mmat{M}^{ab}_{(2)}$, we get $\tilde{\Phi}^a = (\Phi^a)^T$ and
$\tilde{\Phi}^b = (\Phi^b)^T$.

To summarize, the matrix $\Mmat{M}_{(2)}$ is given by
\begin{equation}
 \Mmat{M}_{(2)}=
\begin{pmatrix}
  0 &   \Phi^a : \Theta(\vec{R})  -    [ \Phi^b : \Theta(\vec{R}) ]^T\\
  -\Phi^b : \Theta(\vec{R}) + [ \Phi^a : \Theta(\vec{R}) ]^T & 0\\
\end{pmatrix}.
\label{eq:Mtensor}
\end{equation}
This results is valid for a general pair of objects. If the two objects
are identical, the off-diagonal blocks have the same form with opposite signs.
The additional condition that the entire $\Mmat{M}$ must be symmetrical implies
then that each block by itself is antisymmetric.

By separating the tensors $\Phi$ and $\Theta$ into their symmetric and
antisymmetric parts, the second-order term of the pair-mobility matrix
can be simplified further.
It should be mentioned, in addition, that
the $\Phi$ tensor depends on the origin selected for the object. These
two technical issues are addressed in Appendices
\ref{sec:general_form} and \ref{sec:Phi}, respectively.
Finally, we note that the terms in these tensors corresponding to the 
translational response vanish for spheres and ellipsoids.
Consequently, two such regular objects develop relative velocity only to orders $1/R^3$
and above. 

\section{Numerical analysis for stokeslet objects}
\label{sec:methods}

In the preceding sections we have derived the general properties of
the instantaneous hydrodynamic interaction between two arbitrarily shaped objects.
We now move on to the second part of the work, addressing the time evolution
of the two objects. This complicated problem is not tractable analytically, and we
resort to numerical integration of specific examples.
Because of the complexity of the problem, and since we are interested in generic properties,
we allow ourselves to restrict the analysis to the simplest, even if unrealistic, objects.
Arguably the
simplest form of an arbitrarily shaped object is the so-called {\it
  stokeslet object}\,---\,a discrete set of small spheres, separated
by much larger, rigid distances, where each sphere is approximated as
a point force. 
The sparseness of these objects makes them free-draining, which may be valid for
macromolecules but not for compact objects.

\begin{figure}
\centerline{\resizebox{0.45\textwidth}{!}{\includegraphics[viewport=55 14 345 292]{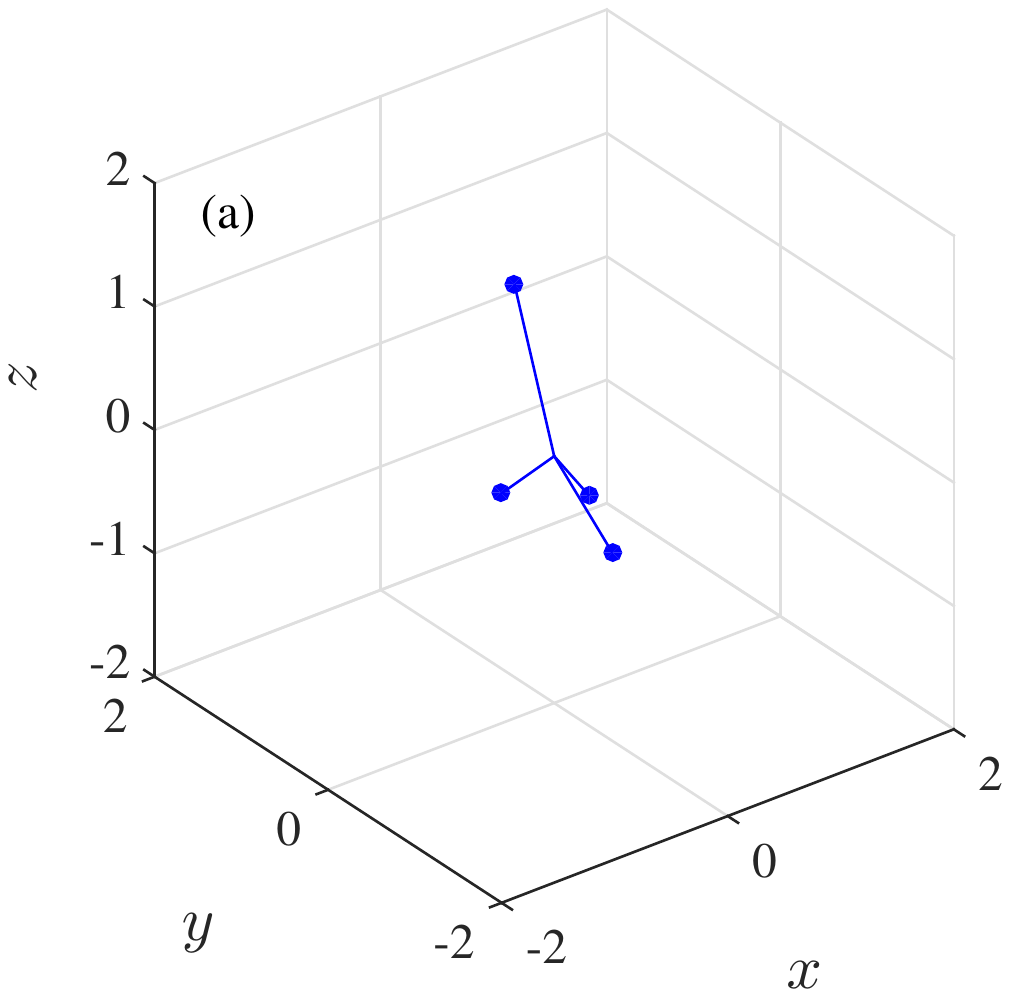}}
\hspace{2cm}
\resizebox{0.45\textwidth}{!}{\includegraphics[viewport=55 14 345 292]{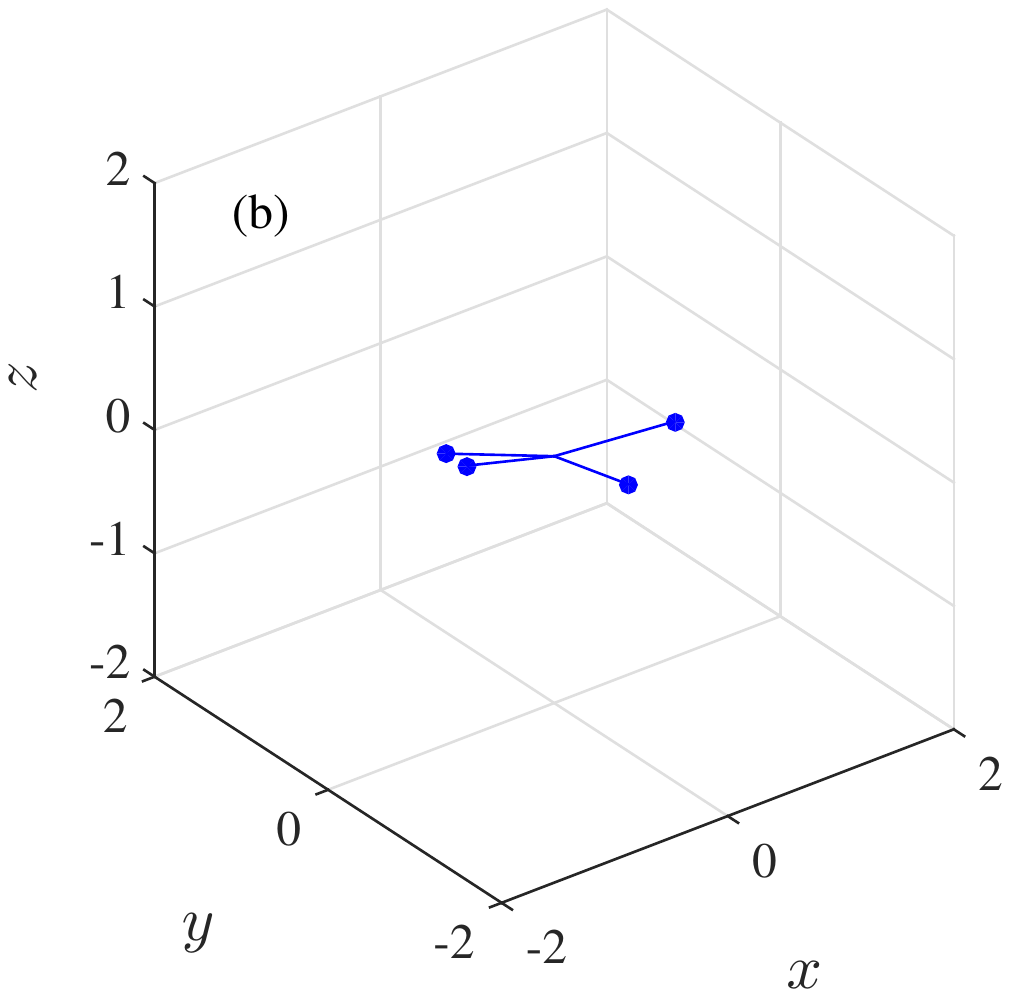}}}
\caption[]{Two examples of axially alignable stokeslet objects, which were used in the simulations. 
The objects comprise four stokeslets connected by dragless rods. The origin of the objects is 
at point (0,0,0) and the aligning direction is $-\hat{z}$.
The object on the left corresponds to the dark red/dotted trajectories in the left panels of
Figs.~\ref{fig:geometry:const} and~\ref{fig:m:const}, and
the one on the right corresponds to the purple/dashed trajectories in the right panels of
Figs.~\ref{fig:geometry:t} and~\ref{fig:m:t}.}
\label{fig:objects}
\end{figure}

We treat pairs of identical objects, each made of four
stokeslets. 
To obtain representative sampling of numerical examples 
we do not design these objects but create them randomly.
Four points are placed at random distances ranging between $0$ and
$1$ from an arbitrary origin. The origin is then shifted to the points'
center of mass.
The radius $\rho$ of the stokeslets is taken as $0.01$. The
resulting configuration is checked to be ``sufficiently chiral'', in
the sense that the $\Mmat{T}$-matrix of the individual object is strongly
asymmetric, having a single real eigenvalue of absolute value
$|\lambda_3|>0.005$, which makes the object axially alignable. (See
Sec.\ \ref{sec:intro}.). Examples of the stokeslet objects
we use are provided in Fig.~\ref{fig:objects}.

The way to calculate the mobility of a single stokeslet object was
presented in Ref.\ \citenum{Krapf_etal2009}. First, we briefly present in
Sec.\ \ref{sec:stokeslet_properties} the simple extension of this
method to pair-mobilities. We calculate both the pair mobility and the
tensor $\Phi$ introduced in Secs.\ \ref{sec:general} and \ref{sec:multipole}.
The latter allows us to
calculate pair mobilities up to second order in the multipole
expansion. Section \ref{sec:numerical} describes how we use the pair
mobility to numerically calculate the time evolution of the pair
configuration.

\subsection{Pair-Mobility and $\Phi$ Tensor}
\label{sec:stokeslet_properties}

The properties of a stokeslet object can be derived self-consistently
from the linear relations which describe the stokeslets' configuration.
This is done without finding the stokeslets' strengths explicitly.  Below
we find the pair-mobility matrix, and the $\Phi$ tensor associated
with a single object, given the stokeslet configuration and the size
of the spheres that they represent.
 
Each of the two objects, $x=a,b$, consists of $N_x$ stokeslets,
$\bvec{F}^x=(\vec{F}^{x}_{1},\dots,\vec{F}^{x}_{N_x})$, in a known
configuration,
$\bvec{r}^x=\left(\vec{r}^{x}_{1},\dots,\vec{r}^{x}_{N_x} \right)$.
Here, we use the notation of a bold letter to denote a set of $N$ 3-vectors,
and $\vec{r}^{x}_{n}$ indicates the position 3-vector of the $n$th
stokeslet in object $x$ with respect to the object's origin.
Each
stokeslet is a sphere of radius $\rho$, where $\rho<\min(r^{x}_{1},\dots
,r^{x}_{N_x})$. The boundary conditions at the sphere surface enter only through its
self-mobility coefficient. The velocities of the spheres, $\vec{v}^{x}_{n}$, are
known from the object's linear and angular velocities,
\begin{equation}
	      \begin{pmatrix}
		\bvec{v}^a \\
		\bvec{v}^b 
	      \end{pmatrix} =
				\begin{pmatrix}
		  \Mmat{U}^{a} & 0\\
		   0 & \Mmat{U}^{b} 
		  \end{pmatrix}
			\begin{pmatrix}
		  \Cvvec{V}^a \\
		  \Cvvec{V}^b 
		  \end{pmatrix}, \qquad 
	    \text{with }
	\Mmat{U}^x=  \begin{pmatrix}
	      \Mmat{I}_{3\times3}  , -\vec{r}^{\,x\,\times}_{1}/l \\ 
				\vdots \\
				\Mmat{I}_{3\times3}  , -\vec{r}^{\,x\,\times}_{N_x}/l
	  \end{pmatrix},
	  \quad \text{for } x=a,b,
	\label{eq:U_velocity}
	\end{equation}
where the matrix $\vec{y}^\times$ obtained from the vector $\vec{y}$ is
defined as $(\vec{y}^\times)_{ij}=\epsilon_{ikj}y_k$.
Each stokeslet force is proportional to the relative velocity of the sphere
that it represents, with respect to the flow around it as created by
the other stokeslets. This gives a linear relation between the
stokeslets and the velocities of the spheres~\footnote{More
  explicitly, consider the stokeslet at position $\vec{r}^a_n$. The
  flow at that point which is created by the other stokeslets,
  belonging to the two objects, is $\vec{u}(\vec{r}^a_n)=\Sigma_{m\neq
    n} \Mmat{G}(\vec{r}^{\,a}_{n}-\vec{r}^{\,a}_{m}) \cdot
  \vec{F}^{\,a}_{m} + \Sigma_{m} \Mmat{G}
  (\vec{R}+\vec{r}^{a}_{n}-\vec{r}^{b}_{m}) \cdot \vec{F}^{\,b}_{m}$.
  The stokeslet at that point is proportional to the velocity of the
  sphere relative to the local flow, $\vec{F}^a_{n}=\gamma
  \left(\vec{v}^a_{n}-\vec{u}(\vec{r}^a_n) \right)$. This gives
  Eq.~(\ref{eq:linear_L}).},
\begin{equation}
	      \begin{pmatrix}
		\bvec{v}^a \\
		\bvec{v}^b 
	      \end{pmatrix} =
	      \begin{pmatrix}
		  \Mmat{L}^{aa}	& \Mmat{L}^{ab}\\
		  {\Mmat{L}^{ab}}^{T} & \Mmat{L}^{bb}
	      \end{pmatrix}
	      \begin{pmatrix}
		  \bvec{F}^a \\
		  \bvec{F}^b 
		  \end{pmatrix}, \quad \text{where }
\label{eq:linear_L}			
\end{equation}
\begin{eqnarray}
			(\Mmat{L}^{xx}_{nm})_{ij} &=&
	      \left\{
	      \begin{array}{l l}
		\Mmat{G}_{ij}(\vec{r}^{\,x}_{n}-\vec{r}^{\,x}_{m}) & \quad \text{if } n \neq m \\
		\gamma^{-1}  \delta_{ij} & \quad \text{else } 
	      \end{array}	    
	      \right. \nonumber\\
					(\Mmat{L}^{ab}_{nm})_{ij} &=&
		\Mmat{G}_{ij} (\vec{R}+\vec{r}^{a}_{n}-\vec{r}^{b}_{m}),
		\label{eq:Lblocks}
\end{eqnarray}
and  $\gamma=6\pi\rho/ l$.

First we find the pair-mobility matrix as a generalization
of the analysis in Ref.~\citenum{Krapf_etal2009}.  The sum of the stokeslets and the
corresponding total torque must be equal to the external generalized forces
applied on the objects. In a matrix form we can write
\begin{equation}
	 \begin{pmatrix}
		  \Cvvec{F}^a \\
		  \Cvvec{F}^b 
		  \end{pmatrix}=
			\begin{pmatrix}
		  (\Mmat{U}^{a})^T & 0\\
		   0 & (\Mmat{U}^{b})^T 
		  \end{pmatrix}
	      \begin{pmatrix}
					\bvec{F}^a \\
					\bvec{F}^b 
	      \end{pmatrix} .
\label{eq:U_force}
\end{equation}
Using Eqs.~(\ref{eq:U_velocity}) and (\ref{eq:linear_L}), we have
$$
\begin{pmatrix}
		  \Mmat{U}^{a} & 0\\
		   0 & \Mmat{U}^{b} 
		  \end{pmatrix}^T
			\cdot
\begin{pmatrix}
		  \Mmat{L}^{aa}	& \Mmat{L}^{ab}\\
		  {\Mmat{L}^{ab}}^{T} & \Mmat{L}^{bb}
	      \end{pmatrix}^{-1}
				\cdot
				\begin{pmatrix}
		  \Mmat{U}^{a} & 0\\
		   0 & \Mmat{U}^{b} 
		  \end{pmatrix}
			\cdot
			\begin{pmatrix}
		  \Cvvec{V}^a \\
		  \Cvvec{V}^b 
		  \end{pmatrix}=
			\begin{pmatrix}
		  \Cvvec{F}^a \\
		  \Cvvec{F}^b.
		  \end{pmatrix} 
			$$
From this expression we identify the pair-mobility matrix as			
\begin{equation}
  \Mmat{M}=\left[\begin{pmatrix}
		  \Mmat{U}^{a} & 0\\
		   0 & \Mmat{U}^{b} 
		  \end{pmatrix}^T
			\cdot
\begin{pmatrix}
		  \Mmat{L}^{aa}	& \Mmat{L}^{ab}\\
		  {\Mmat{L}^{ab}}^{T} & \Mmat{L}^{bb}
	      \end{pmatrix}^{-1}
				\cdot
				\begin{pmatrix}
		  \Mmat{U}^{a} & 0\\
		   0 & \Mmat{U}^{b} 
		  \end{pmatrix} \right]^{-1}.
\label{eq:fullM}
\end{equation}
This expression allows to calculate the pair-mobility matrix,
with the help of Eqs. (\ref{eq:U_velocity}) and (\ref{eq:Lblocks}),
based on the stokeslets' configuration and the Oseen tensor alone.

Next, we derive the $\Phi^x$ tensor of a stokeslet object $x$.
From this tensor we may readily obtain the second multipole of the pair
interaction  (cf.\ Sec.\ \ref{sec:multipole}). 
The force dipole around the origin
of a forced object is given by [Eq.\ (\ref{eq:V2a})], $(\bvec{r}\bvec{F})^x
\equiv (\Phi^x)^T \cdot \Cvvec{F}^x$. Similar to the $\Mmat{U}^x$
matrix relating the stokeslets to the total generalized force,
$\Cvvec{F}^x=(\Mmat{U}^x)^T \cdot \bvec{F}^x$, we define a tensor of
rank 3, $\Upsilon^x$, which relates the stokeslet forces to the total force
dipole on the object by $(\bvec{r}\bvec{F})^x=(\Upsilon^x)^T \cdot
\bvec{F}^x$. (Note that no force dipole
is applied on the individual stokeslets; being arbitrarily small they possess only 
a force monopole.) 
Specifically, it is made of $N$ blocks of $3\times
3\times 3$, given by $(\Upsilon_n)_{ijs} = r_{n,s}\delta_{ij}$,
$n=1\dots N$, $i,j,s=1,2,3$ (i.e., $r_{n,s}$ is the $s$ Cartesian
coordinate of the stokeslet $n$).  Using Eqs.~(\ref{eq:U_velocity})
and (\ref{eq:linear_L}), we have $(\bvec{r}\bvec{F})^a=(\Upsilon^a)^T
\cdot (\Mmat{L}^{aa})^{-1} \cdot \Mmat{U}^a \cdot \Cvvec{V}^a$. This
implies $(\Phi^x)^T = (\Upsilon^x)^T \cdot (\Mmat{L}^{xx})^{-1} \cdot
\Mmat{U}^x \cdot \Mmat{M}_{\text{self}}^x$. Recalling that the
matrices $\Mmat{M}_{\text{self}}$ and $\Mmat{L}$ are symmetric, we
  finally get 
\begin{equation}
  \Phi^x = \Mmat{M}_{\text{self}}^x \cdot
  (\Mmat{U}^x)^T \cdot (\Mmat{L}^{xx})^{-1} \cdot \Upsilon^x.
\label{eq:stokeslets_Phi}
\end{equation}

It is important to note that in the above derivation we compute
$\Mmat{M}$ and $\Phi$ under the assumption that, for each object, the
stokeslet sizes are arbitrarily small compared to the distances
between them, $\rho \ll l$ 
(where $l$ is the object's radius of gyration).
However, in a more general case one can
use the Rotne-Prager-Yamakawa
tensor~\cite{Rotne&Prager1969,Yamakawa1970}, which corrects the Oseen
tensor for force distributions with finite
size~\cite{Carrasco&Torre1999}.

\subsection{Numerical Time Integration}
\label{sec:numerical}

We present a numerical integration scheme for the dynamics of two
stokeslet objects. We should first define the reference frames used in
the scheme. Each rigid object is characterized by axes which are
affixed and rotate with it. We define the object reference frame (ORF)
such that its $z$ axis coincides with the object's alignment axis (the
corresponding eigenvector of the $\Mmat{T}$-matrix). The other two axes
are selected arbitrarily. The $z$ axis of the laboratory frame is
defined along the forcing direction. During the evolution we
follow the translation and rotation of the ORF in the laboratory
frame.

We represent the orientation of an object by the Euler{-}Rodrigues
4-parameters \cite{Favro1960}, defined by
$(\Gamma,\vec{\Omega}) \equiv
(\cos\frac{\theta}{2},\hat{n}\sin\frac{\theta}{2})$, where $\hat{n}$
and $\theta$ are the axis and angle of rotation \footnote{This is the
  same as the unit-quaternion representation \cite{Favro1960}.}. The
following properties hold for this 4-parameter representation: (a) The
norm of $(\Gamma,\vec{\Omega})$ in 4D-space is unity,
$\Gamma^2+\Omega^2=1$.  (b) A rotation matrix is given by Rodrigues'
rotation formula,
\begin{equation}
R(\Gamma,\vec{\Omega})= \Mmat{I}_{3 \times 3} +2\Gamma \vec{\Omega}^{\times}+
2 ( \vec{\Omega}^{\times} )^2.
\end{equation}
(c) The parameters are invariant under inversion, i.e.,
$(\Gamma,\vec{\Omega})$ and $(-\Gamma,-\vec{\Omega})$ correspond to
the same orientation.  (d) Given the angular velocity of the object,
the dynamics of its orientation simply reads
	\begin{equation}
		\begin{pmatrix}
		\dot{\Gamma} \\
		\dot{\vec{\Omega}}
	\end{pmatrix}=
	\frac{1}{2}
	\begin{pmatrix}
		0  & -\vec{\omega}^T\\
		\vec{\omega} &\vec{\omega}^{\times}
	\end{pmatrix}
	\begin{pmatrix}
		\Gamma \\
		\vec{\Omega}
	\end{pmatrix}.
\label{eq:angular}
	\end{equation}

Since we choose the ORF such that the $z$-axis is the axis of
alignment, the terminal orientation of an axially alignable object under
a constant force along the $z$-axis is
$(\Gamma,\vec{\Omega})=(\cos(\frac{\omega t + \alpha}{2} ),\hat{z}
\sin(\frac{\omega t+ \alpha}{2}) )$, where $\alpha$ is a constant phase
which depends on the object's initial orientation at time $t=0$.

The state of a pair of objects at time $t$ is described by the
position of the origins of the objects, $\vec{R}^a(t)$ and
$\vec{R}^b(t)$, and their orientation parameters, $(
\Gamma^a(t),\vec{\Omega}^a(t) )$ and $( \Gamma^b(t),\vec{\Omega}^b(t)
)$.  We time-integrate from the initial state,
$\vec{R}^a_{0}=(0,0,0)$, $\vec{R}^a_0-\vec{R}^b_{0}=\vec{R}_0$, $(
\Gamma^a_{0},\vec{\Omega}^a_{0} )$ and $(
\Gamma^b_{0},\vec{\Omega}^b_{0} )$, as follows.  Given the positions
of the stokeslets at time $t$, the pair-mobility matrix,
$\Mmat{M}(t)$, is calculated as explained in
Sec.\ \ref{sec:stokeslet_properties}, either exactly or using the
multipole approximation.  Then, the linear and angular velocities of
the objects are given by $ (\Cvvec{V}^a (t),\Cvvec{V}^b (t))^T =
\Mmat{M}(t) \cdot (\Cvvec{F}^a(t),\Cvvec{F}^b (t) )^T$. From them
the origins and orientations of the objects at time $t+dt$ are derived
according to
\begin{eqnarray}
  \vec{R}^x(t+dt)&=&\vec{R}^x(t)+\vec{V}^x(t)dt
\\
\begin{pmatrix}
 \Gamma^x(t+dt) \\
 \vec{\Omega}^x(t+dt)
\end{pmatrix} &=& 
\exp \left[ 	\frac{dt}{2}
	\begin{pmatrix}
		0  & -\vec{\omega}^{xT}\\
		\vec{\omega}^x &\vec{\omega}^{x\times}
	\end{pmatrix}
\right]
\begin{pmatrix}
      \Gamma^x(t) \\
      \vec{\Omega}^x(t)
\end{pmatrix}
\end{eqnarray}
for $x=a,b$.  During the evolution we make sure that the small
stokeslet spheres do not overlap, and that the pair mobility matrix
remains positive-definite.  In practice we never encountered such
problems when using the exact pair mobility matrices; when it did
happen in the case of the multipole approximation we stopped the
integration.
 
We define a scalar order parameter which characterizes the degree of
mutual alignment of the two objects,
\begin{equation}
 m \equiv \left[ ( \Gamma^a,\vec{\Omega}^a) \cdot (
   \Gamma^b,\vec{\Omega}^b) \right]^2= \left( \Gamma^a\Gamma^b +
 \vec{\Omega}^a \cdot \vec{\Omega}^b \right)^2.
\label{eq:phi}
\end{equation}
As required, the order parameter is invariant under 3-rotation. This
can be verified by explicitly applying a 3-rotation to the laboratory
frame, or alternatively, by the following argument. Since 3-rotation
leaves the norm of the 4-parameter orientation unchanged (property (a)
above), it is a unitary transformation in 4-space. Hence, the dot
product is invariant. When the objects are aligned, $ (
\Gamma^a,\vec{\Omega}^a)=\pm ( \Gamma^b,\vec{\Omega}^b)$, and $m=1$;
otherwise $0 \leq m <1$. In the case of partial alignment,
$m=\cos^2(\frac{\Delta\alpha}{2})$, where $\Delta\alpha$ is the mutual
phase difference~\footnote{If the symmetry of the objects is such that
  their phase difference is unobservable (e.g., two ellipsoids
  rotating around their major axis), then we set it to zero.}.

Another scalar property of the two-object system is the energy dissipation rate.
At time $t$, the latter is given by
$\Cvvec{V}^a(t)\cdot\Cvvec{F}^a(t)+\Cvvec{V}^b(t)\cdot\Cvvec{F}^b(t)$.
Since the pair-mobility matrix is positive definite the energy dissipation of
the driven pair is positive at all times.

\section{Numerical Results: Effect on Alignment} 
\label{sec:alignment}

\begin{figure}
\centerline{\resizebox{0.4\textwidth}{!}{\includegraphics[viewport=14 2 389 302]{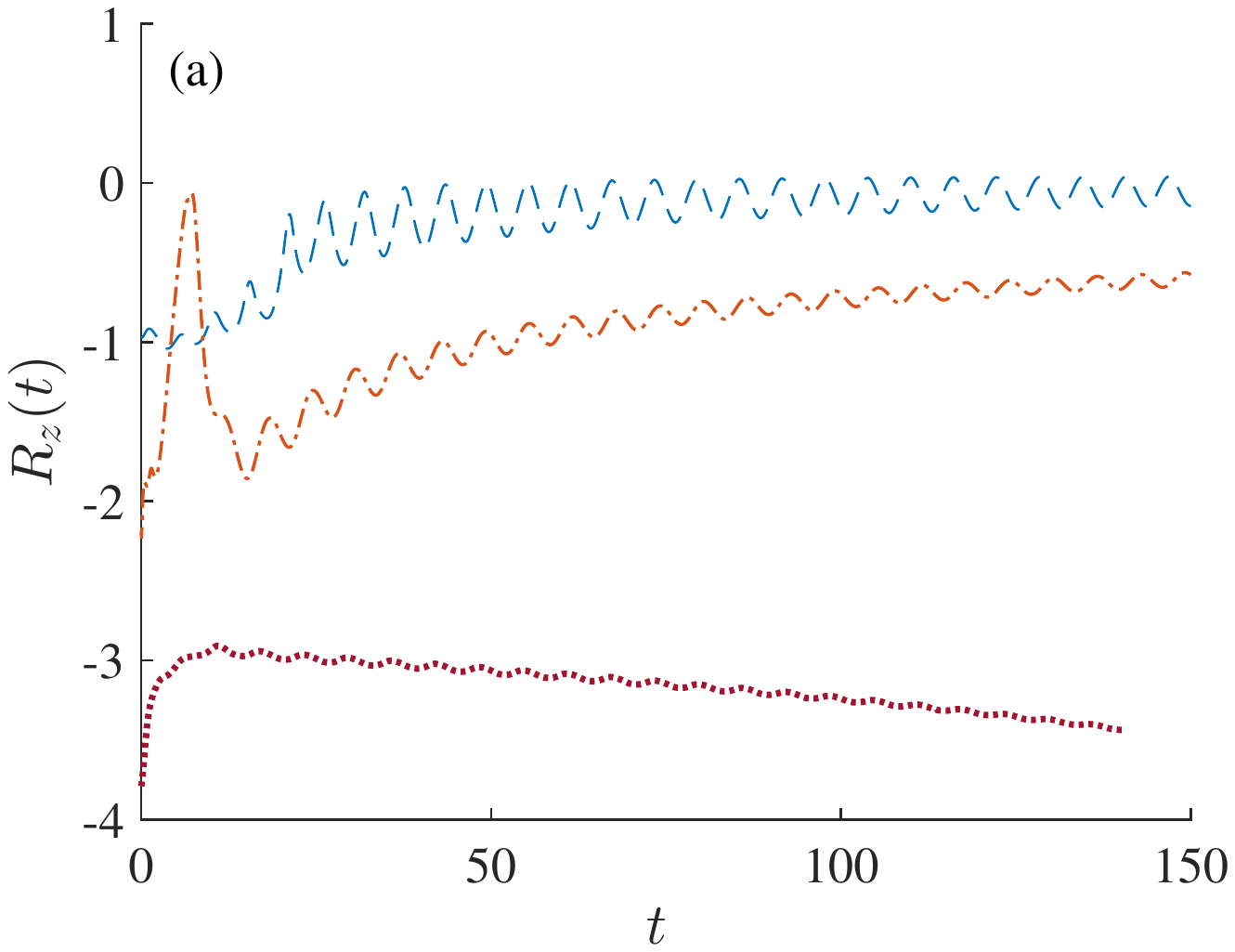}}
\hspace{0.55cm}
\resizebox{0.4\textwidth}{!}{\includegraphics[viewport=14 2 380 304]{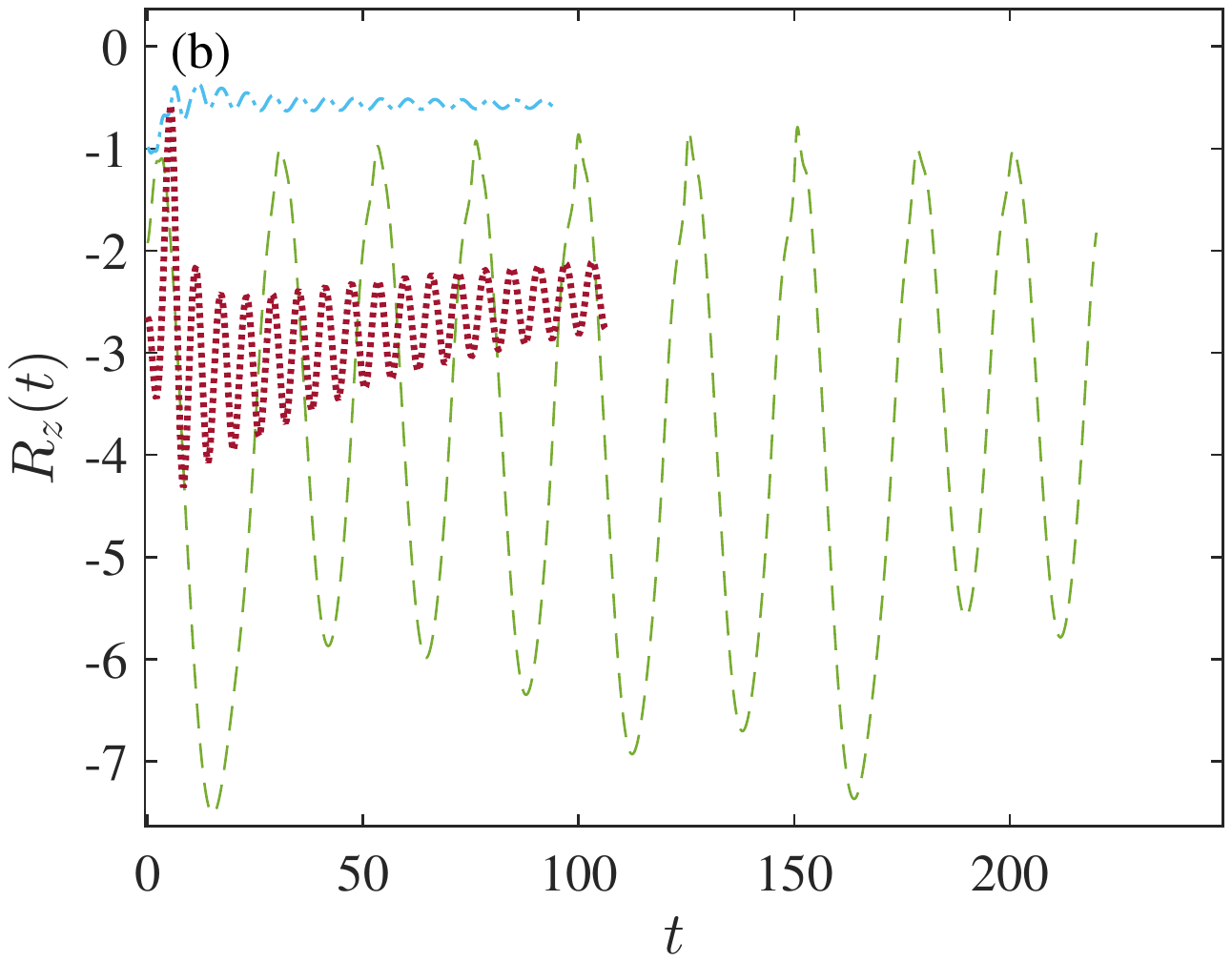}}}
\vspace{0.3cm}
\centerline{\resizebox{0.4\textwidth}{!}{\includegraphics[viewport=51 0 346 301]{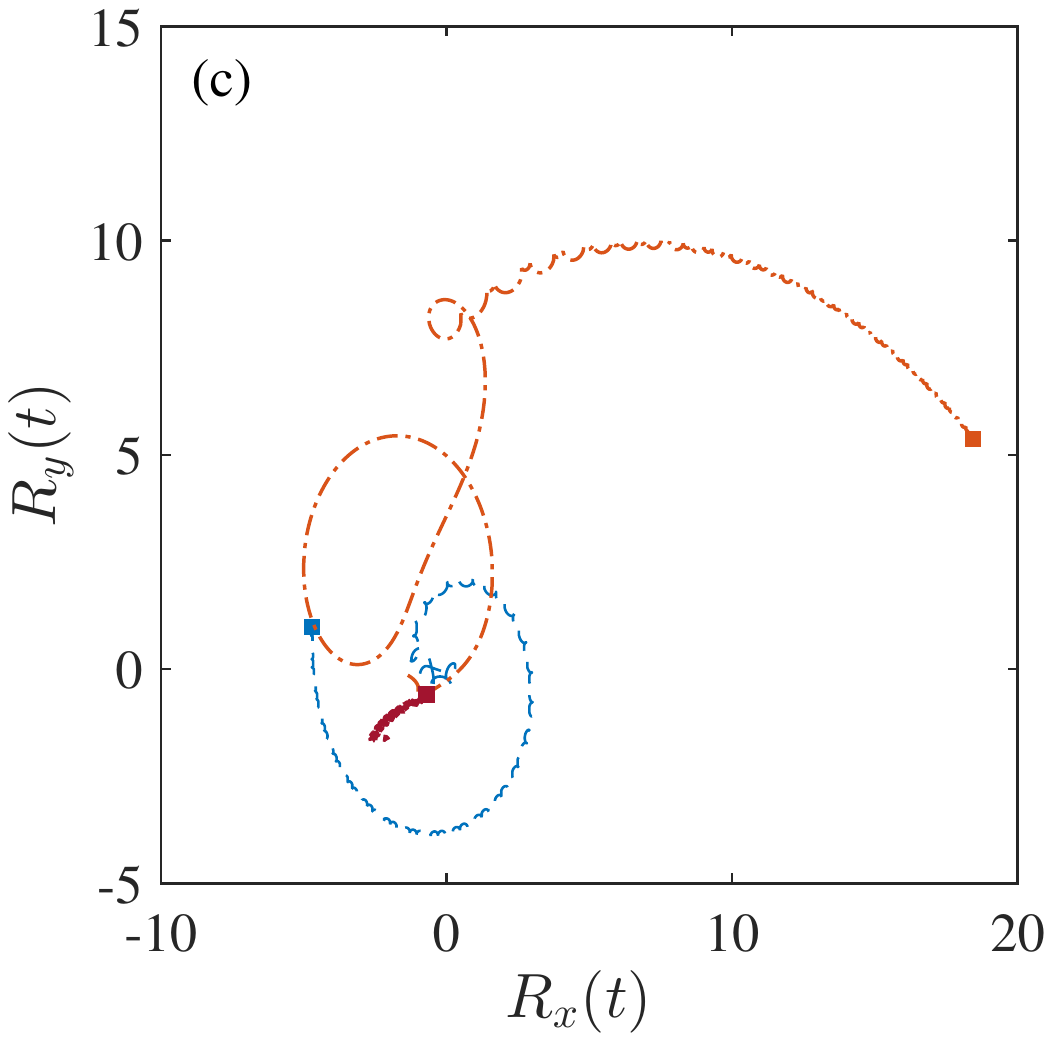}}
\hspace{0.55cm}
\resizebox{0.4\textwidth}{!}{\includegraphics[viewport=45 0 346 301]{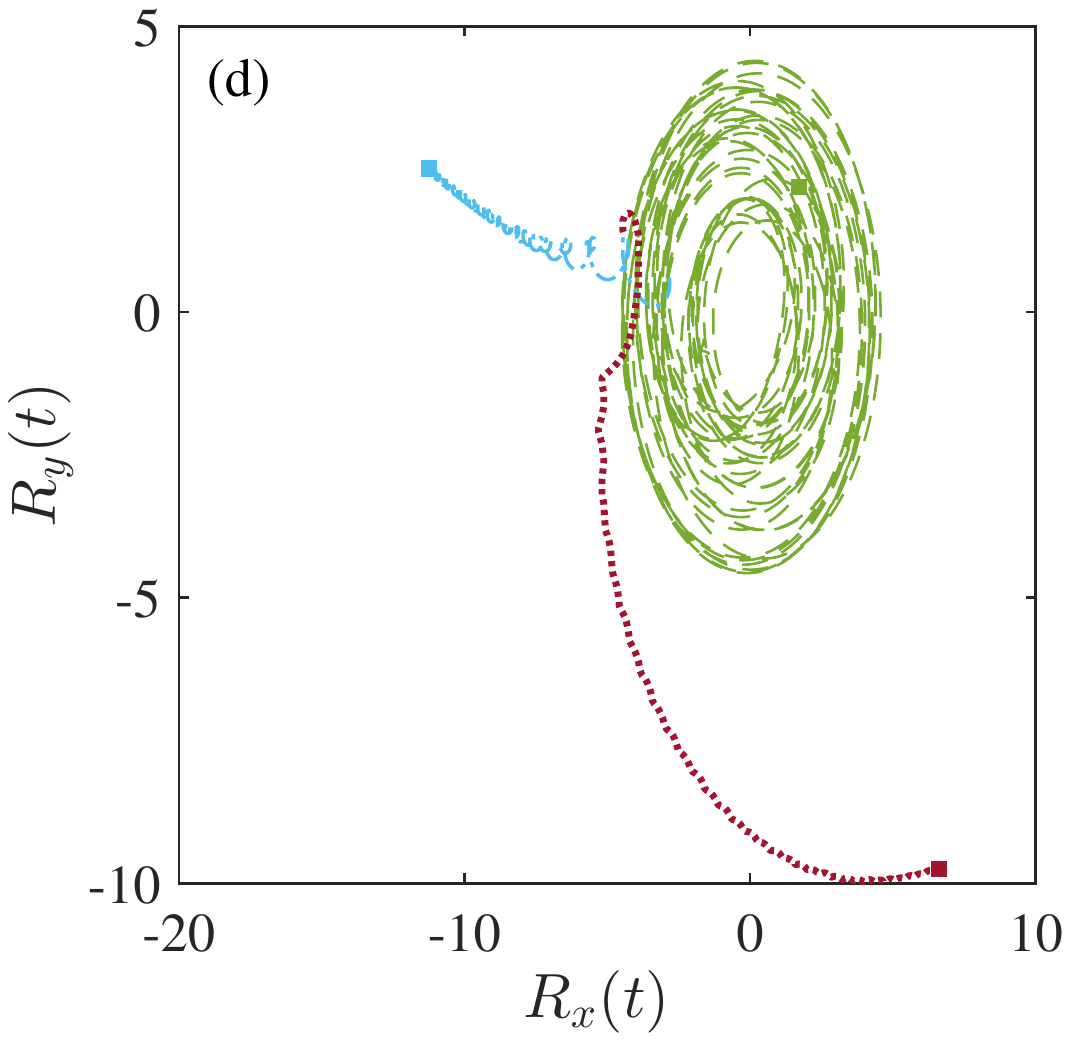}}}
\vspace{0.3cm}
\centerline{\resizebox{0.4\textwidth}{!}{\includegraphics[viewport=11 2 389 302]{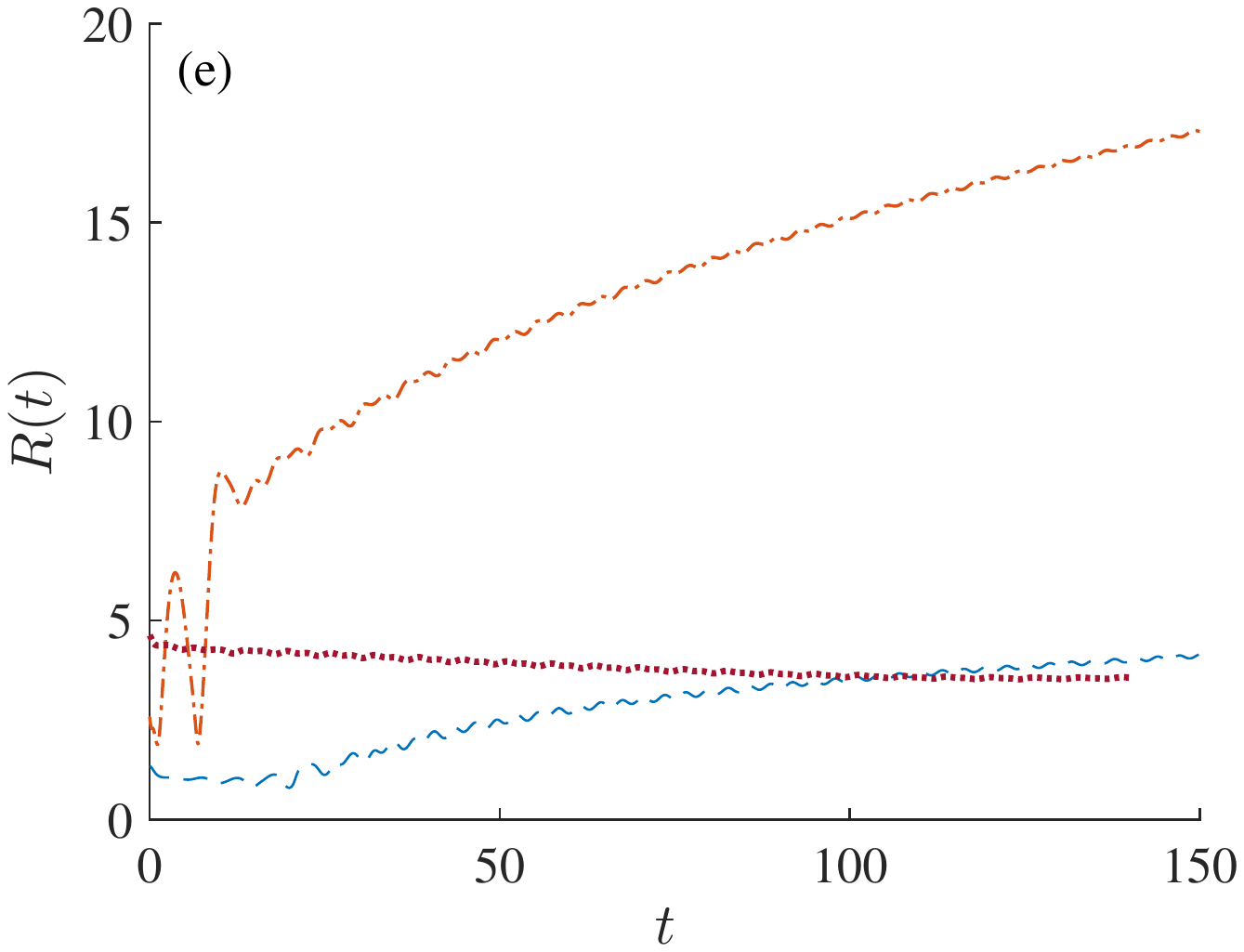}}
\hspace{0.55cm}
\resizebox{0.4\textwidth}{!}{\includegraphics[viewport=13 4 388 302]{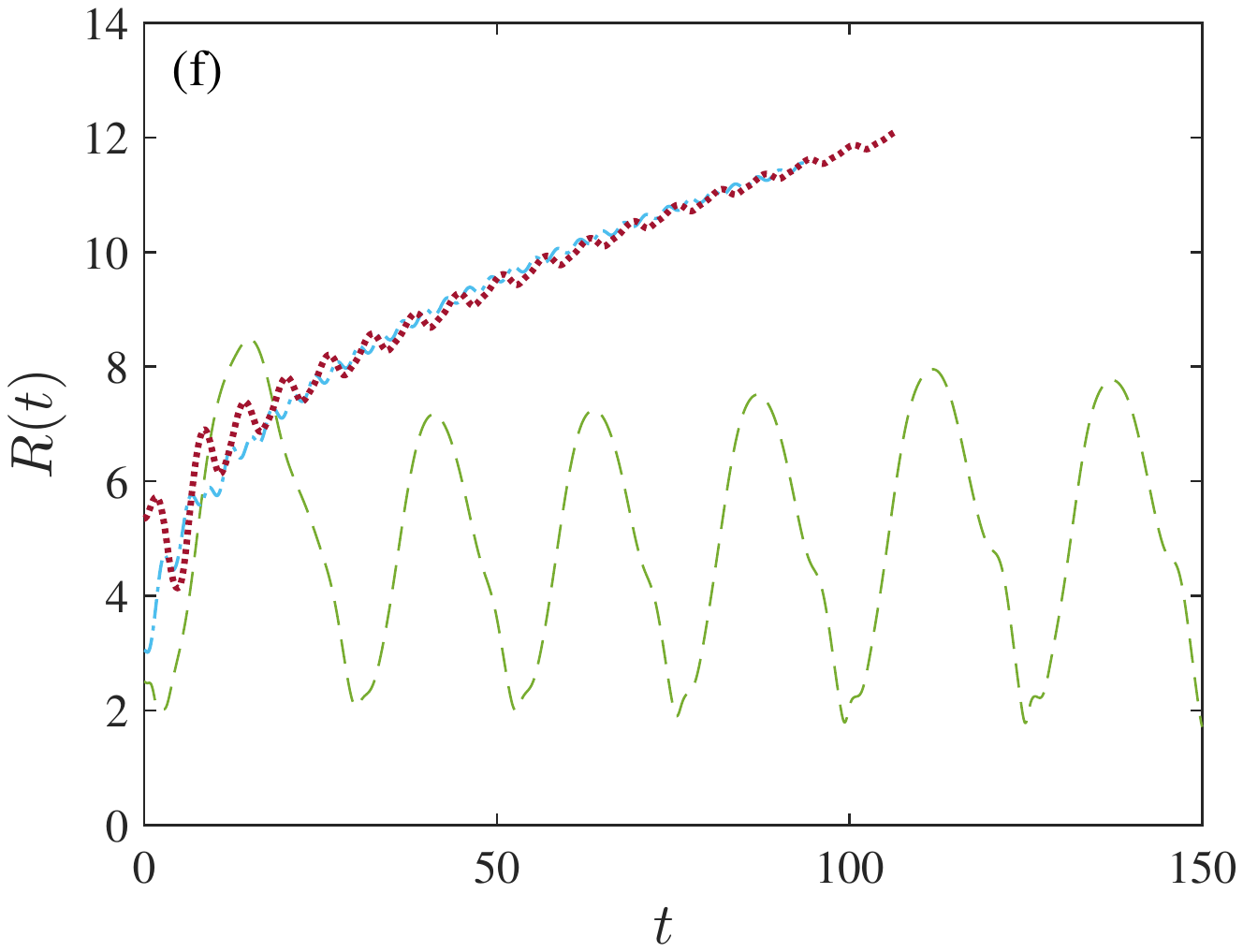}}}
\caption[]{Trajectories of object separation under time-dependent
  forcing. The three rows, from top to bottom, correspond,
  respectively, to the separation along the $z$ direction, its
  projection onto the $xy$ plane, and its total magnitude. The squares
  in the middle row indicate the state at the end of the
  simulation. The panels show results for three different
  objects, starting from either a random mutual orientation (left
  column) or their fully aligned state (right column).
	The green/dashed trajectory on the right panels was integrated
	longer than 150 time units to verify that it continues in a limit cycle.}
\label{fig:geometry:t}
\end{figure}

\begin{figure}
\centerline{\resizebox{0.45\textwidth}{!}{\includegraphics[viewport=7 2 390 300]{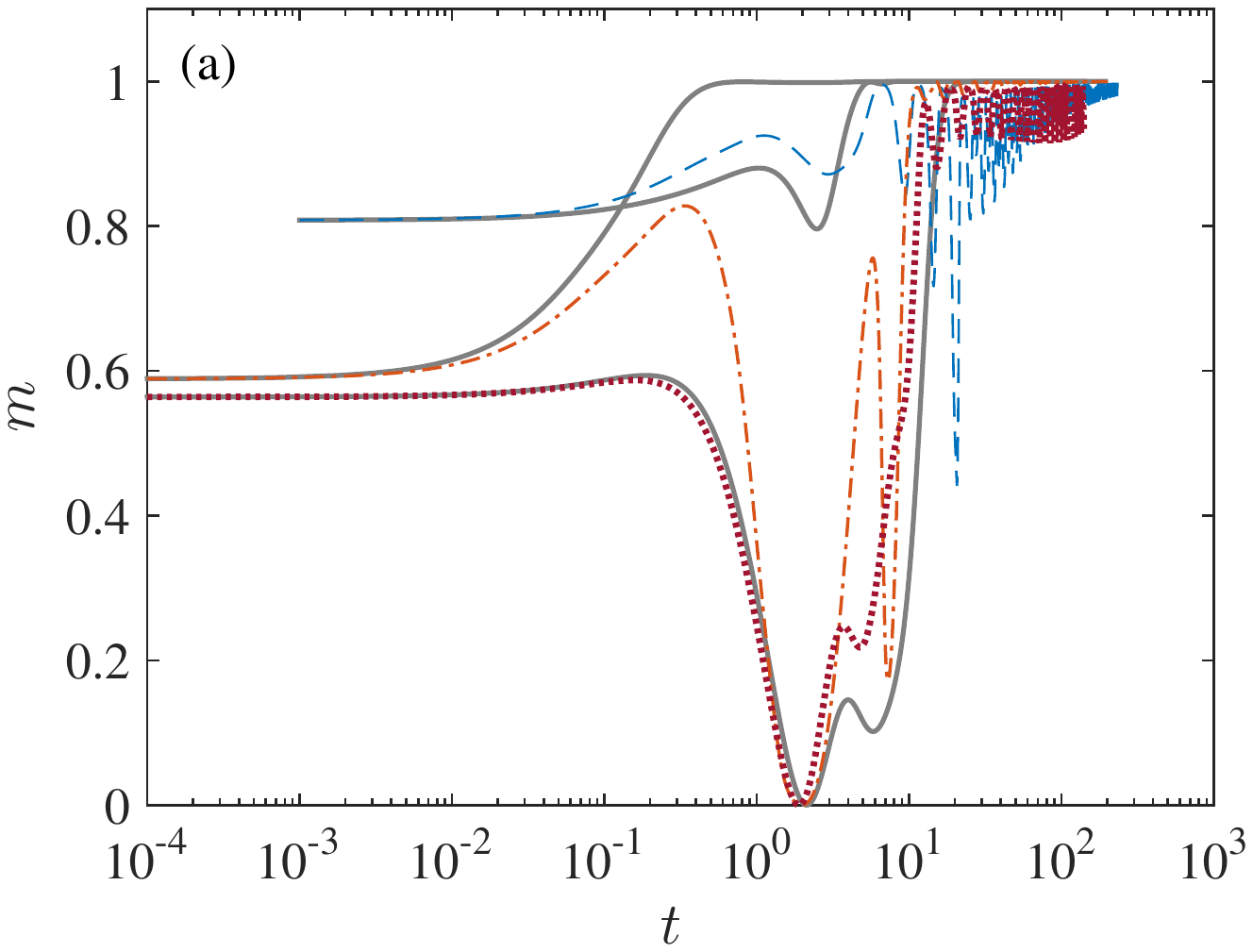}}
\hspace{0.35cm}
\resizebox{0.45\textwidth}{!}{\includegraphics[viewport=7 2 380 302]{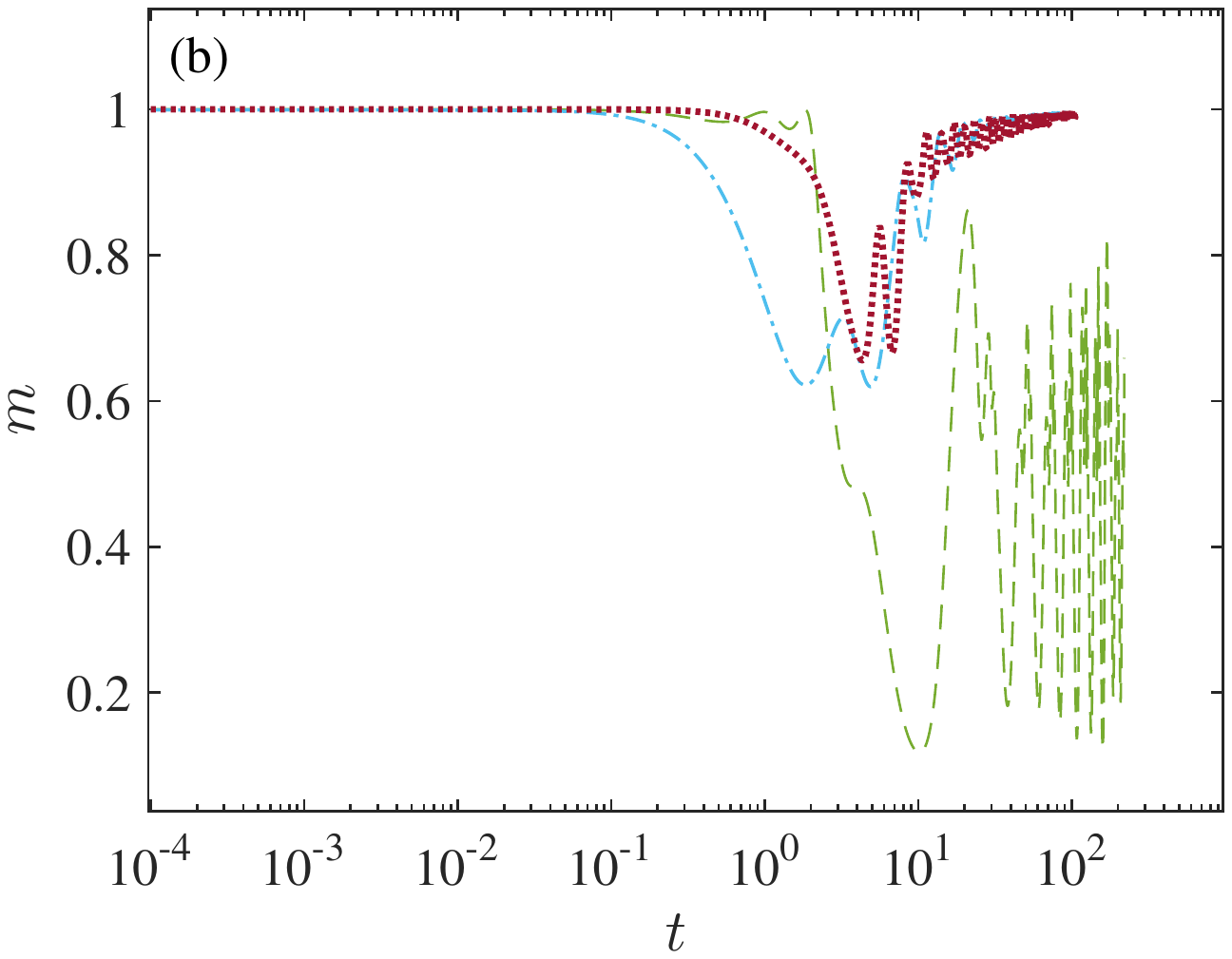}}}
\caption[]{Orientation order parameter as a function of time, under
  time-dependent forcing, for the examples of
  Fig.\ \ref{fig:geometry:t}. (a) results for
  random initial orientations (examples on the left column of
  Fig.\ \ref{fig:geometry:t}); the additional solid gray curves
  correspond to non-interacting objects. (b) 
	results for initially fully aligned object pairs (right column in
  Fig.\ \ref{fig:geometry:t}).}
\label{fig:m:t}
\end{figure}

\begin{figure}
\centerline{\resizebox{0.45\textwidth}{!}{\includegraphics[viewport=14 2 389 302]{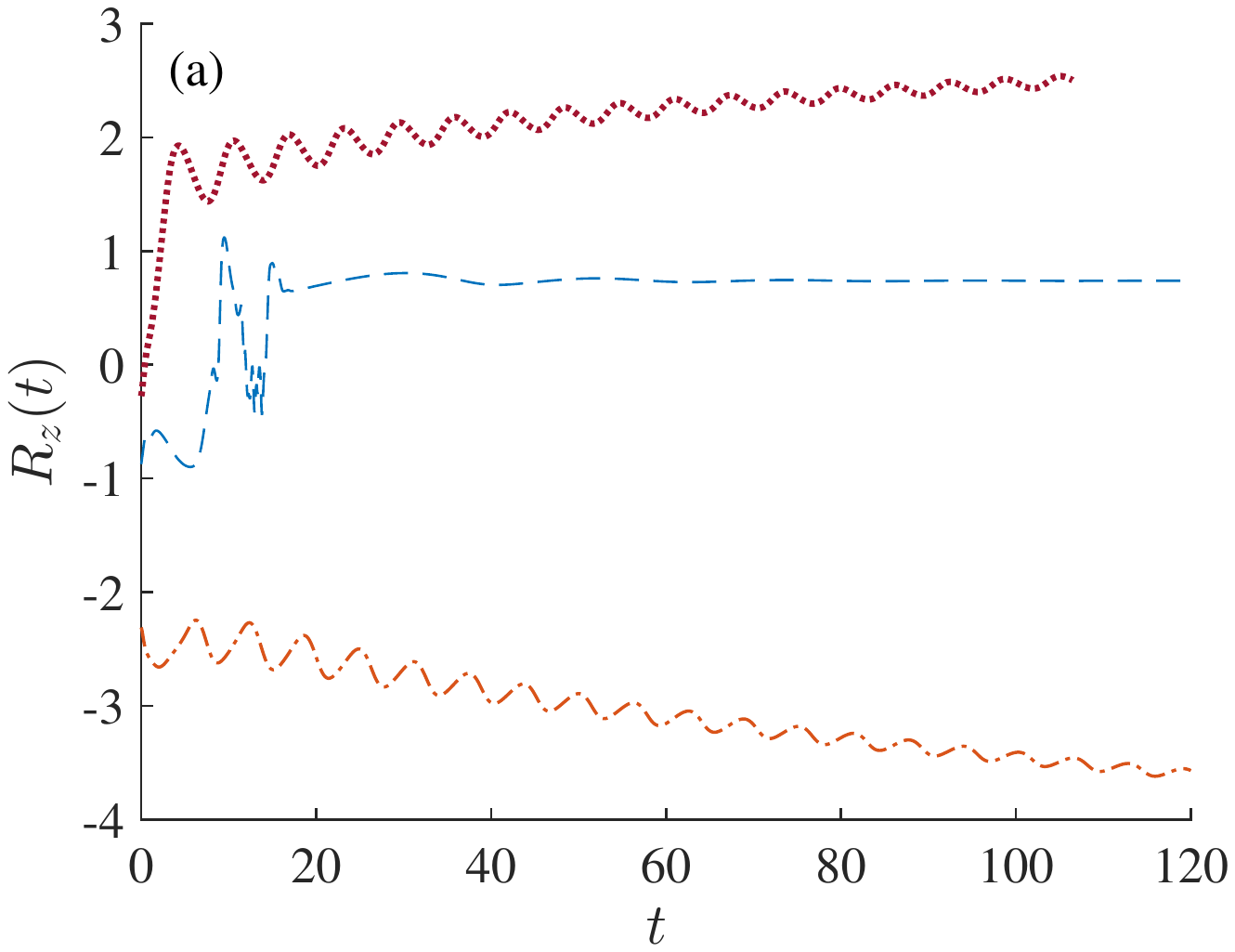}}
\hspace{0.55cm}
\resizebox{0.45\textwidth}{!}{\includegraphics[viewport=14 2 389 304]{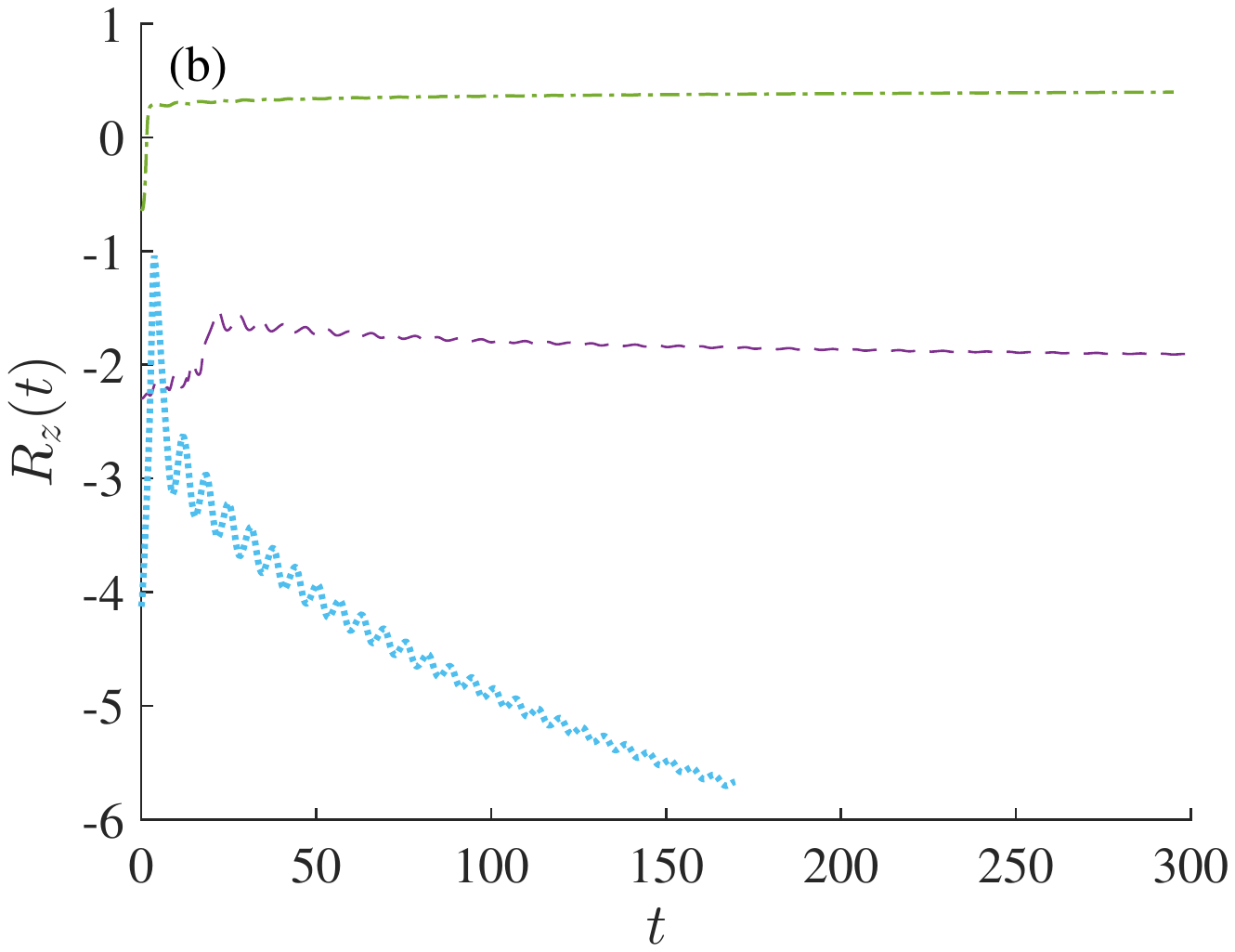}}}
\vspace{0.3cm}
\centerline{\resizebox{0.45\textwidth}{!}{\includegraphics[viewport=45 0 343 301]{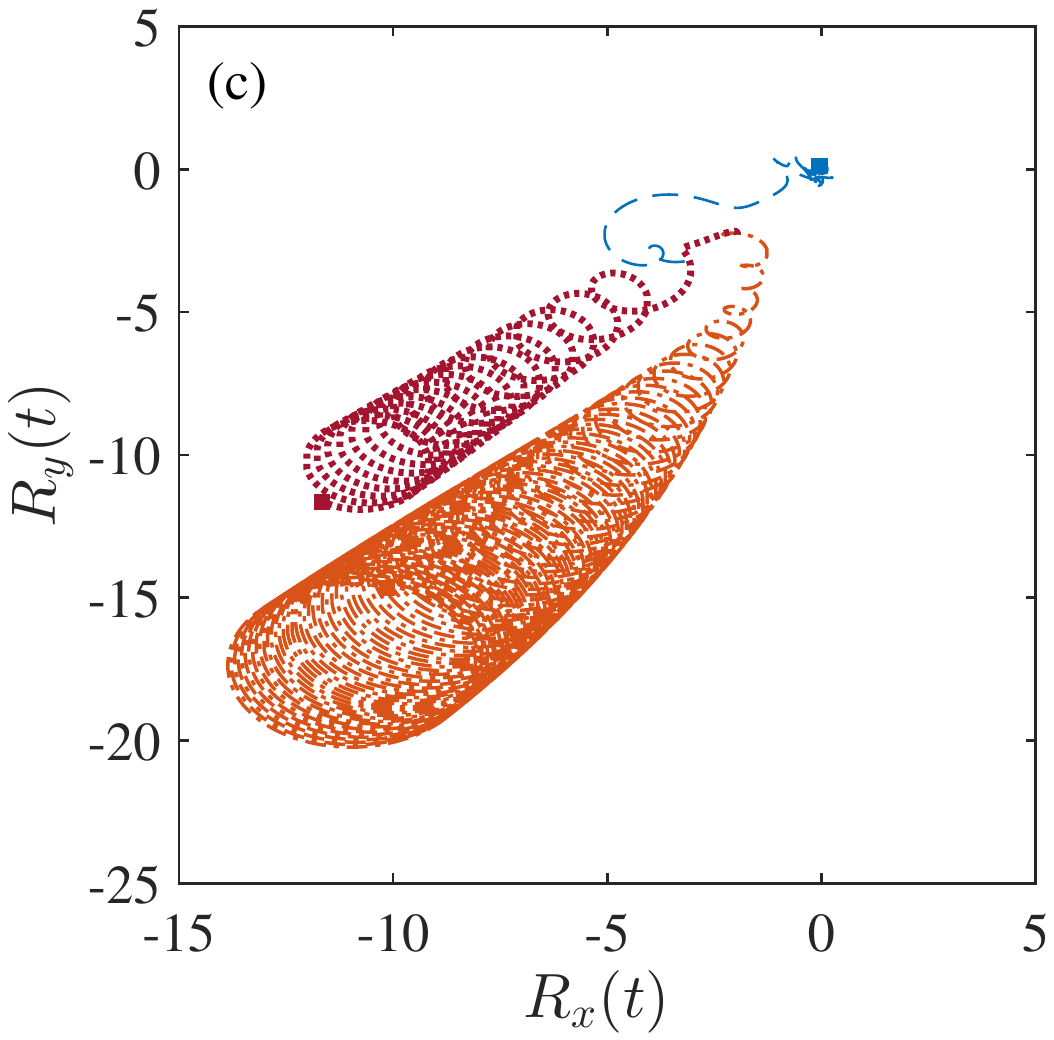}}
\hspace{0.55cm}
\resizebox{0.45\textwidth}{!}{\includegraphics[viewport=45 0 346 301]{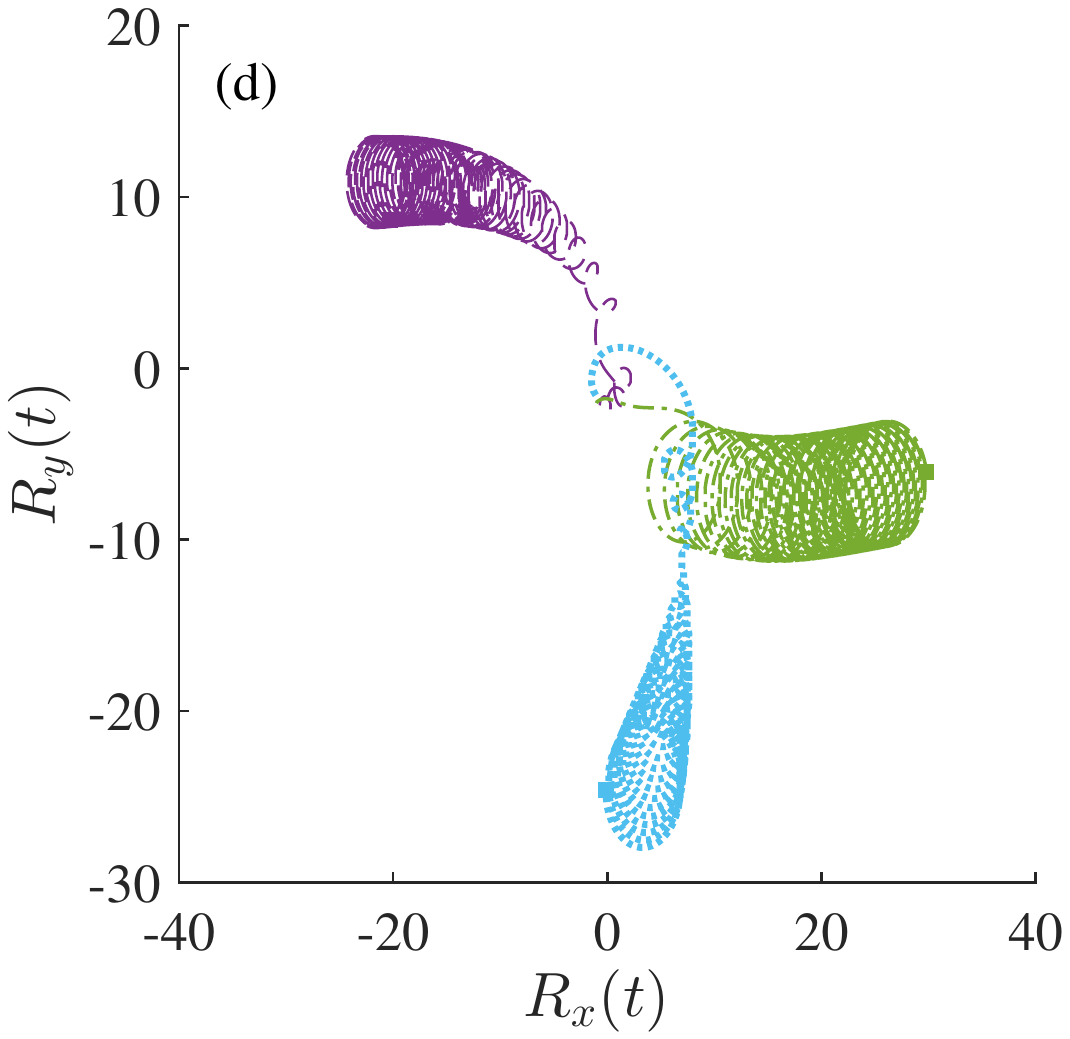}}}
\vspace{0.3cm}
\centerline{\resizebox{0.45\textwidth}{!}{\includegraphics[viewport=11 2 390 303]{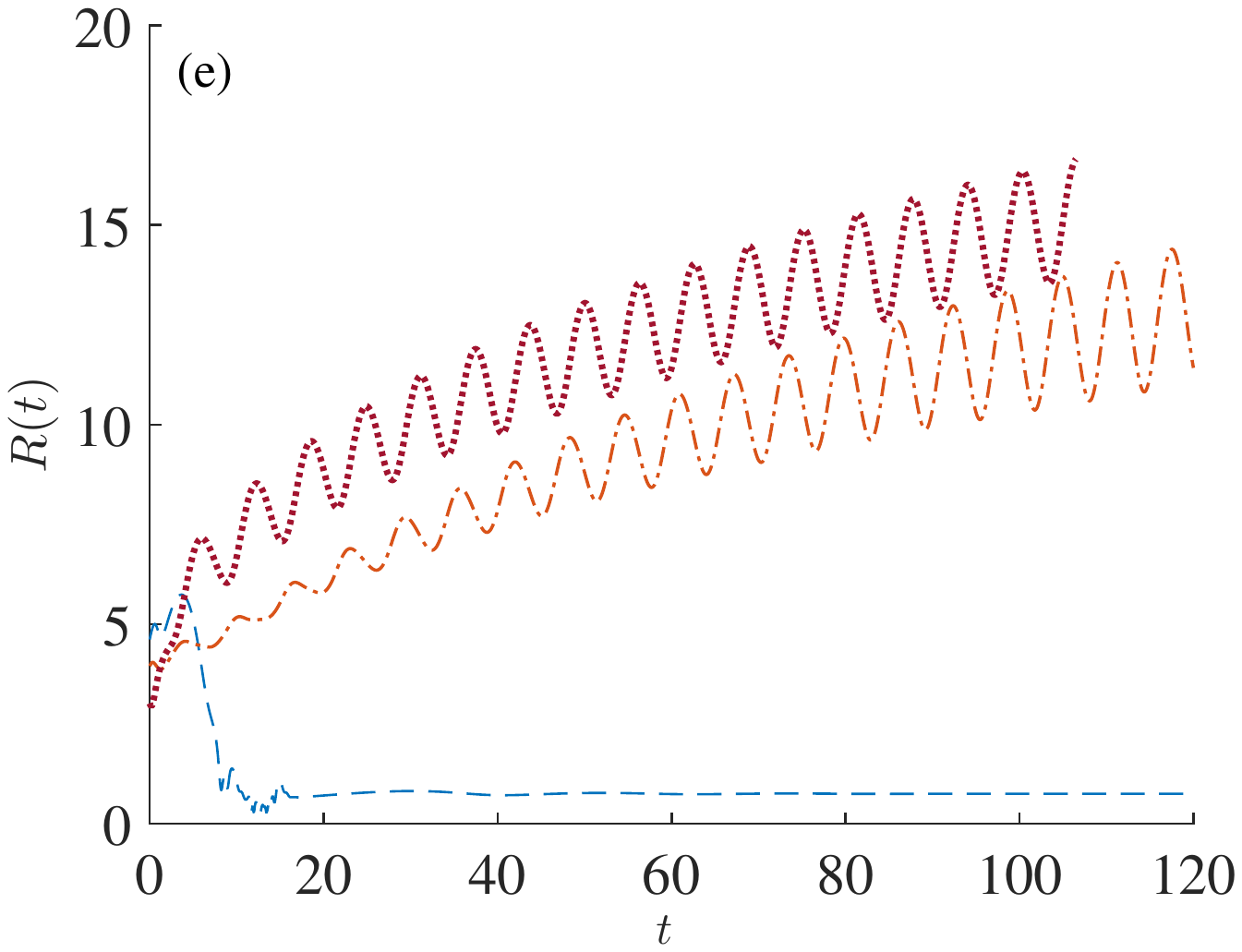}}
\hspace{0.55cm}
\resizebox{0.45\textwidth}{!}{\includegraphics[viewport=13 4 389 302]{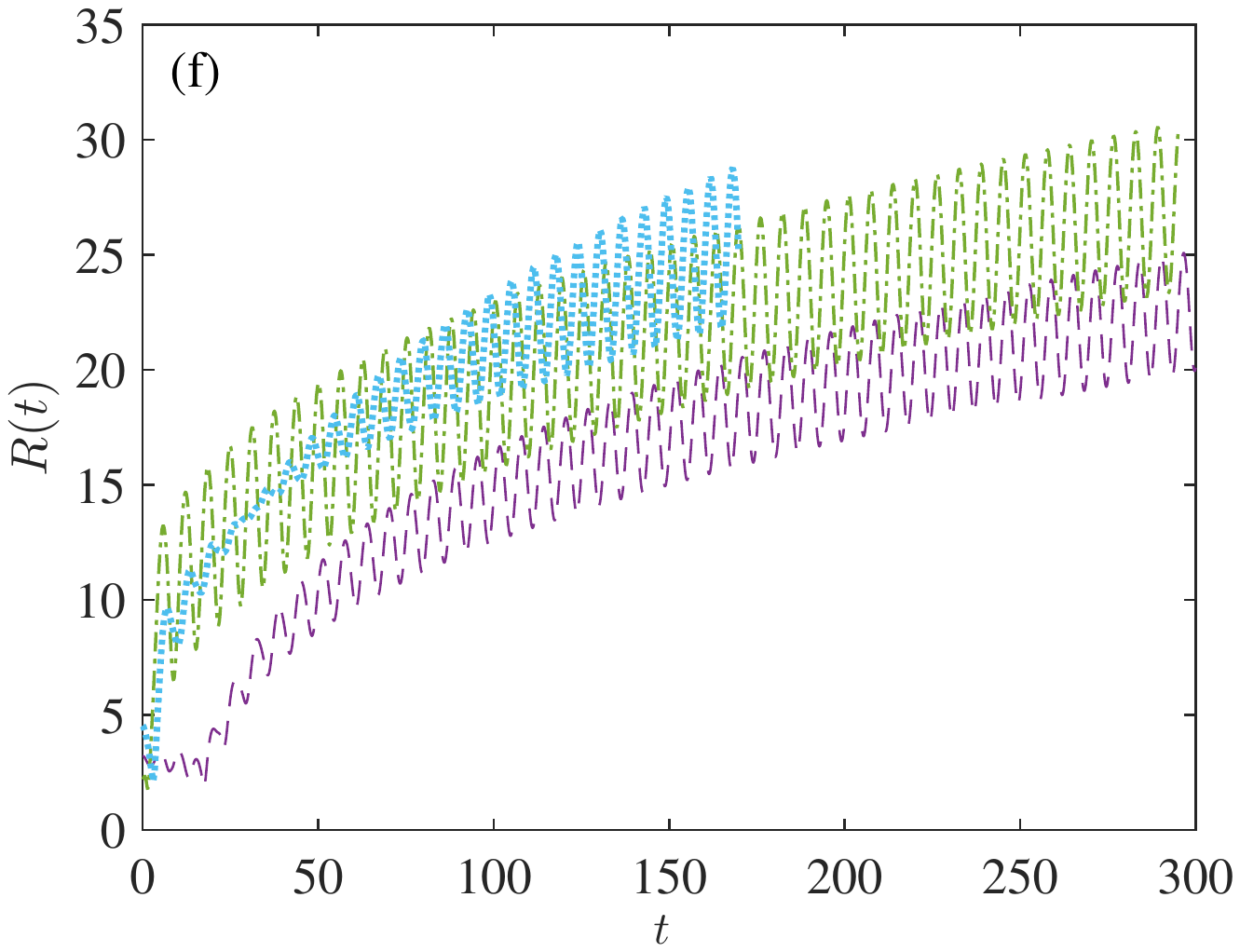}}}
\caption[]{Trajectories of particle separation under constant
  forcing. The meaning of the various panels is the same as in
  Fig.\ \ref{fig:geometry:t}. 
	In all the examples shown here, the two objects repel each other except of the example
	which corresponds to the blue/dashed curve in the left panels.
  (The repulsive trajectories were actually integrated to times longer than presented here.).
	}
\label{fig:geometry:const}
\end{figure}

\begin{figure}
\centerline{\resizebox{0.45\textwidth}{!}{\includegraphics[viewport=7 2 380 302]{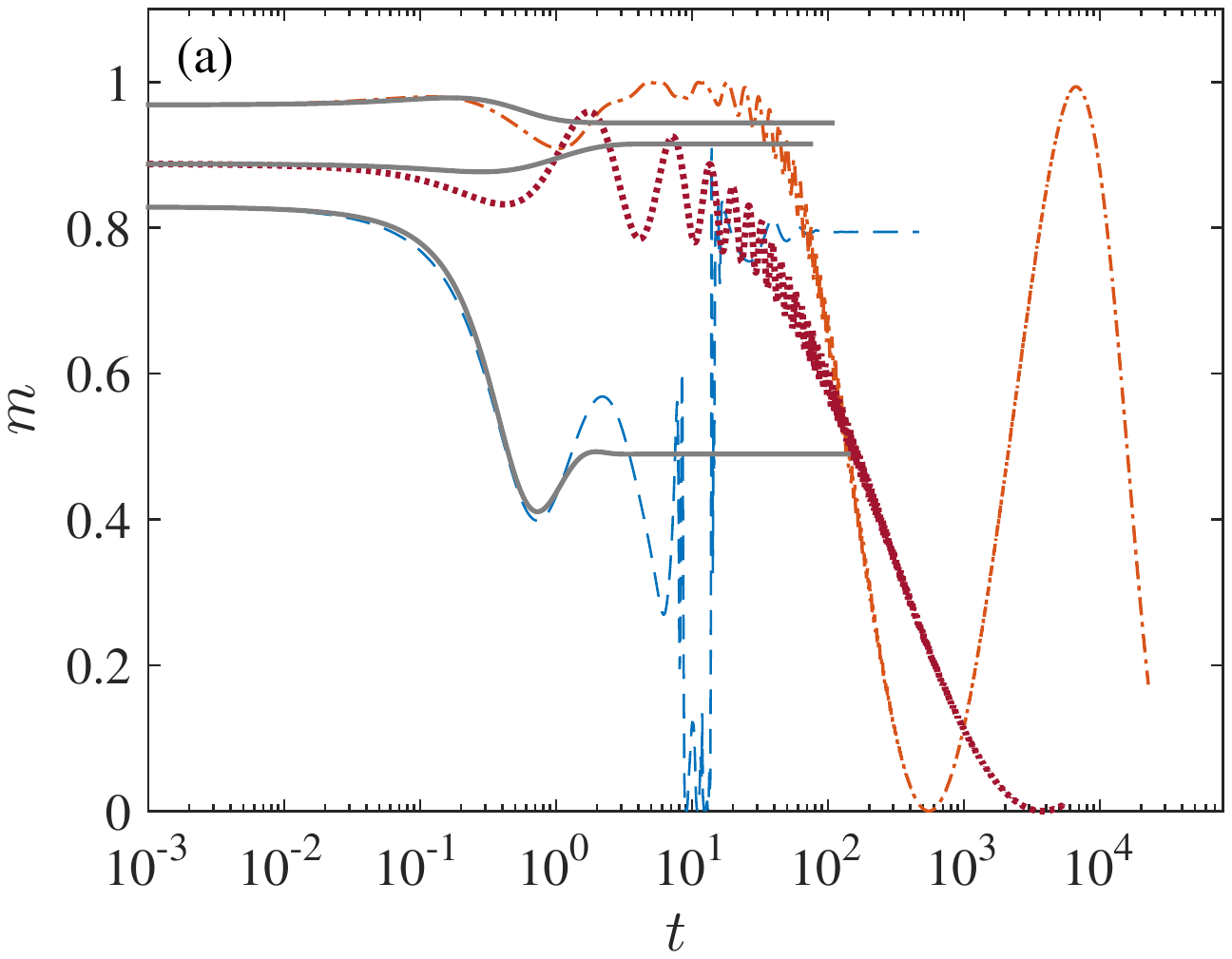}}
\hspace{0.35cm}
\resizebox{0.45\textwidth}{!}{\includegraphics[viewport=7 2 380 305]{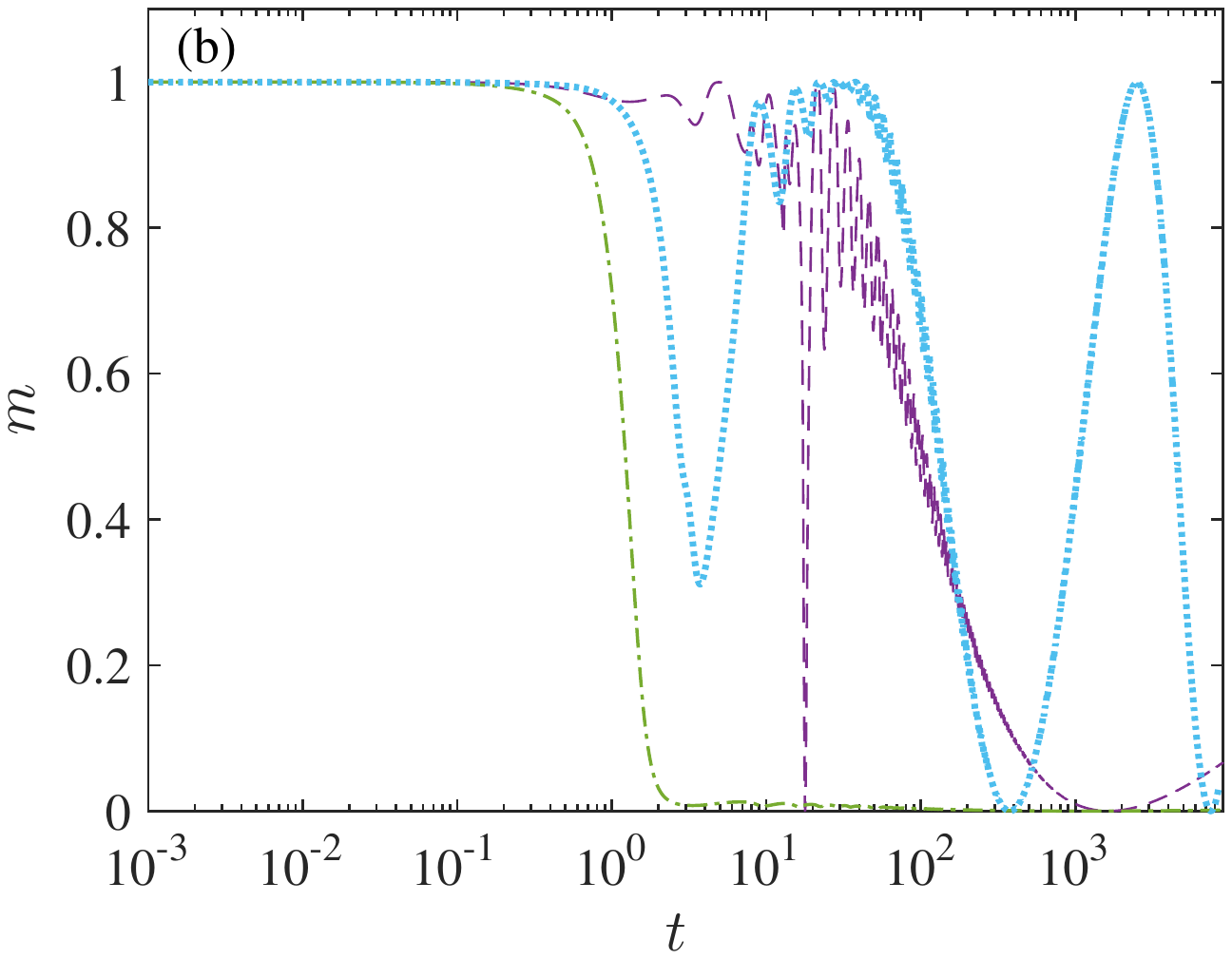}}}
\vspace{0.3cm}
\centerline{\resizebox{0.45\textwidth}{!}{\includegraphics[viewport=7 2 380 302]{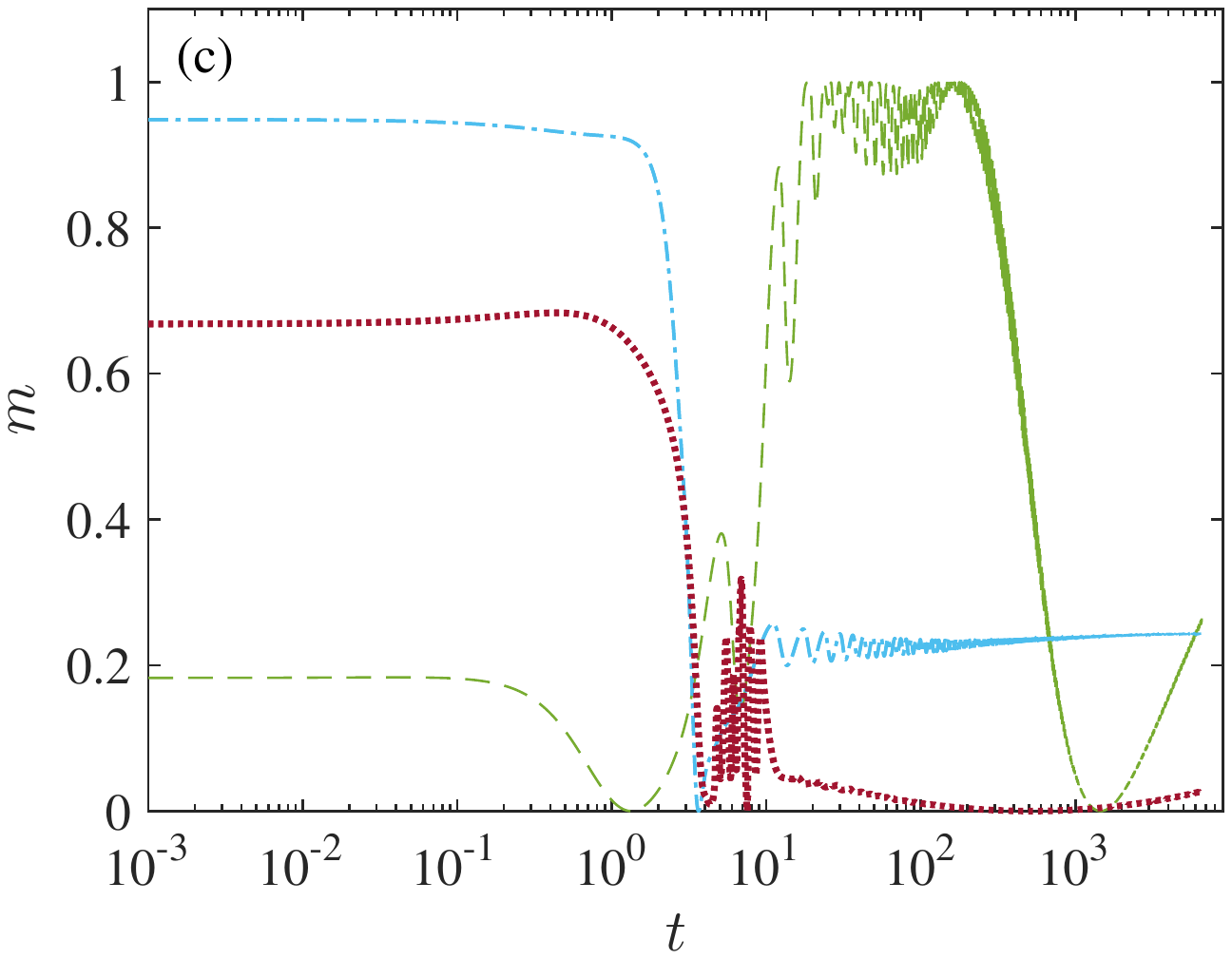}}
\hspace{0.35cm}
\resizebox{0.45\textwidth}{!}{\includegraphics[viewport=7 2 390 301]{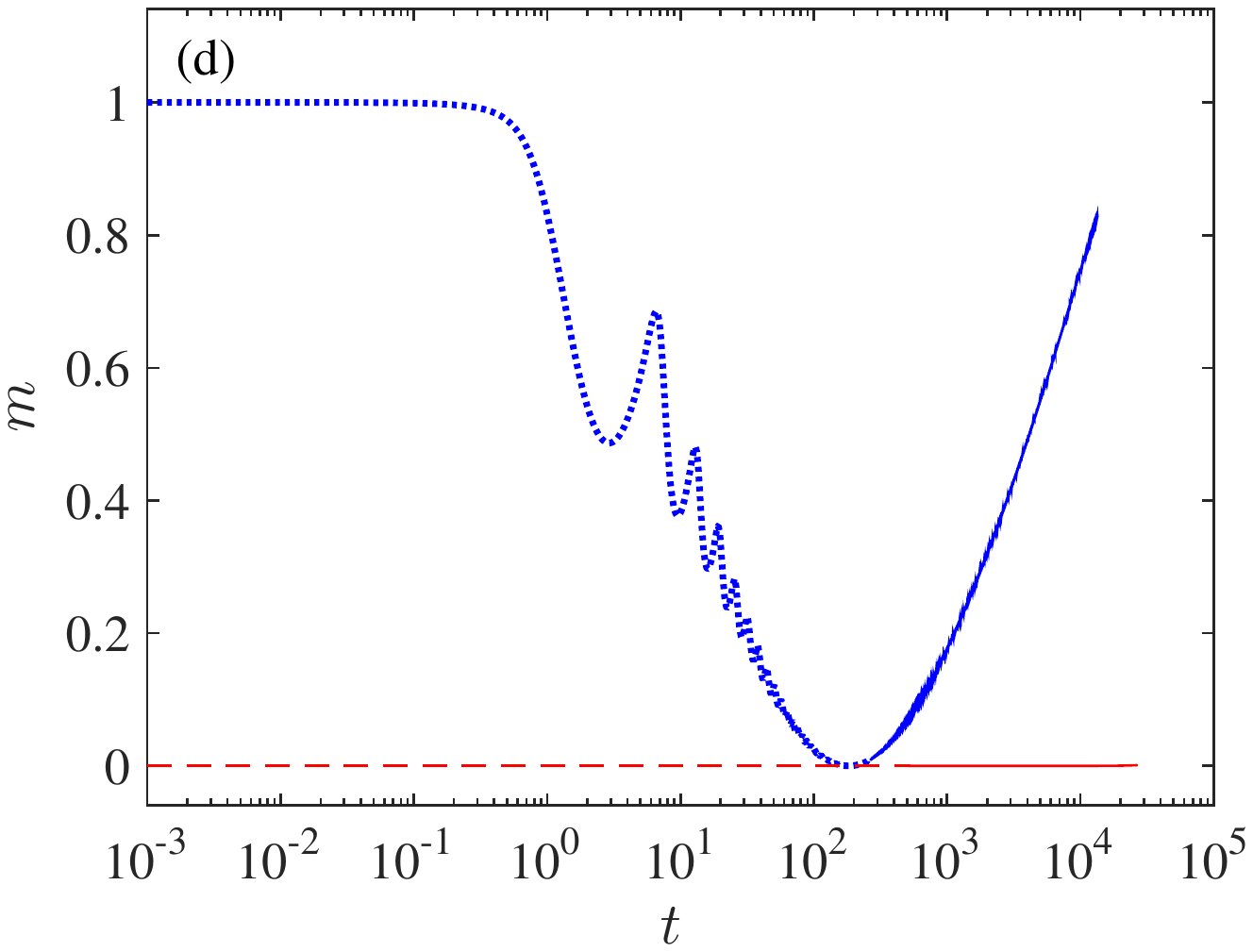}}}
\caption[]{Orientation order parameter as a function of time, under
  constant forcing, for the examples of
  Fig.\ \ref{fig:geometry:const}. 
	(a) results for
  random initial orientations (examples on the left column of
  Fig.\ \ref{fig:geometry:const}); the solid gray curves correspond
  to non-interacting objects. (b) results for
  initially fully aligned object pairs (right column in
  Fig.\ \ref{fig:geometry:t}). (c) 
  results for objects with initial partial alignment (rotating around
  the same axis with random initial phases).
	(d) the stability of anti-alignment;
	shows trajectories of two identical pairs, which start on the
	$xy$ plane from the same separation and axes of rotation but with different relative phases.
	Blue/dotted and red/dashed curves represent, respectively, a pair which starts aligned (zero relative phase)
	and one which starts anti-aligned (relative phase of $\pi$).}
\label{fig:m:const}
\end{figure}

\begin{figure}
\centerline{\resizebox{0.45\textwidth}{!}{\includegraphics[viewport=2 2 380 301]{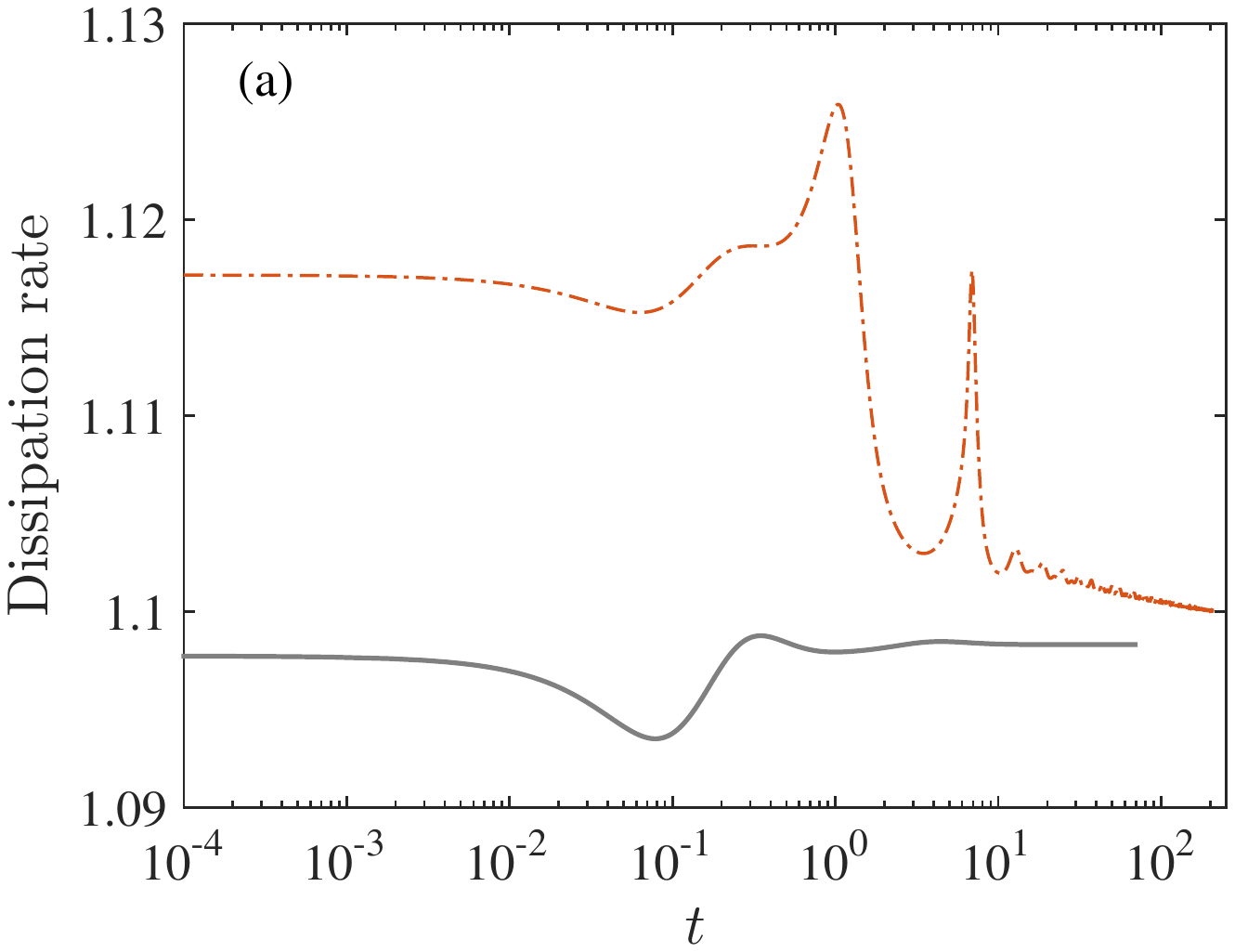}}
\hspace{0.35cm}
\resizebox{0.45\textwidth}{!}{\includegraphics[viewport=2 2 380 301]{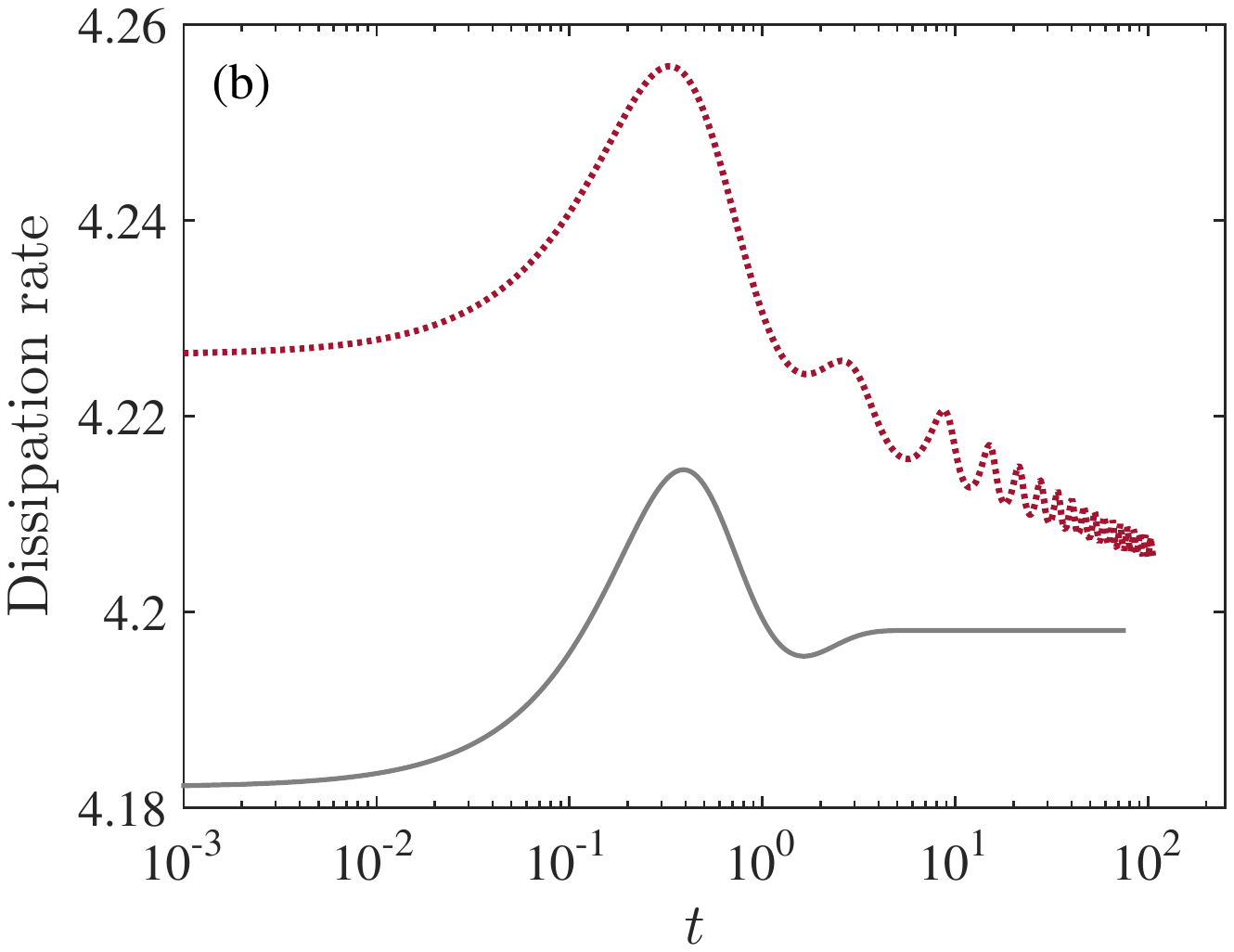}}}
\caption[]{Dissipation rate as a function of time for object pairs
  starting from arbitrary orientations, under time-dependent forcing
  (a) and constant forcing (b). Dash-dotted and dotted colored curves correspond to the
  examples of the same styles/colors in the preceding figures. Solid curves
  show the results in the absence of HI.}
\label{fig:dissip}
\end{figure}

\begin{figure}
\centerline{\resizebox{0.45\textwidth}{!}{\includegraphics[viewport=11 2 389 302]{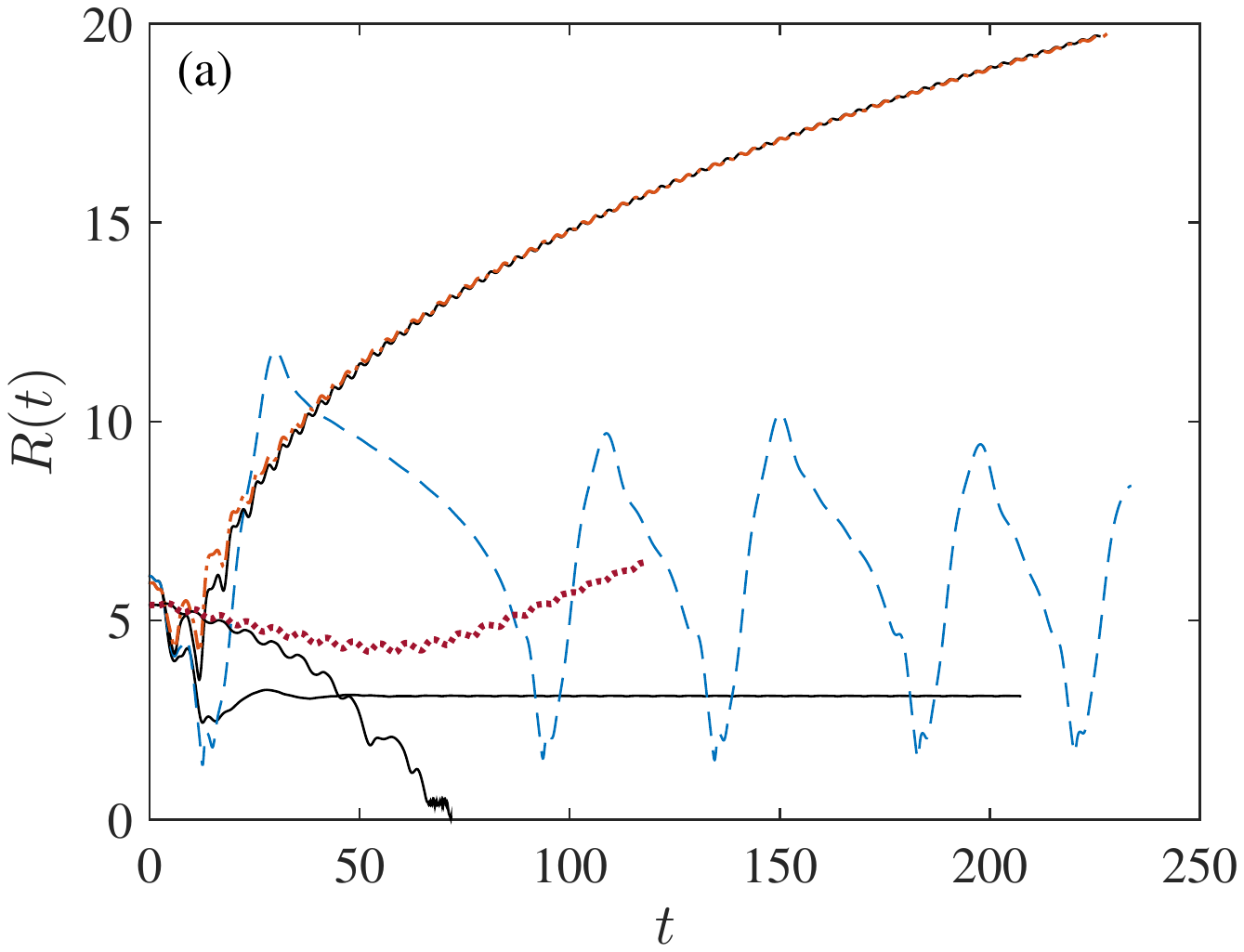}}
\hspace{0.35cm}
\resizebox{0.45\textwidth}{!}{\includegraphics[viewport=11 0 389 302]{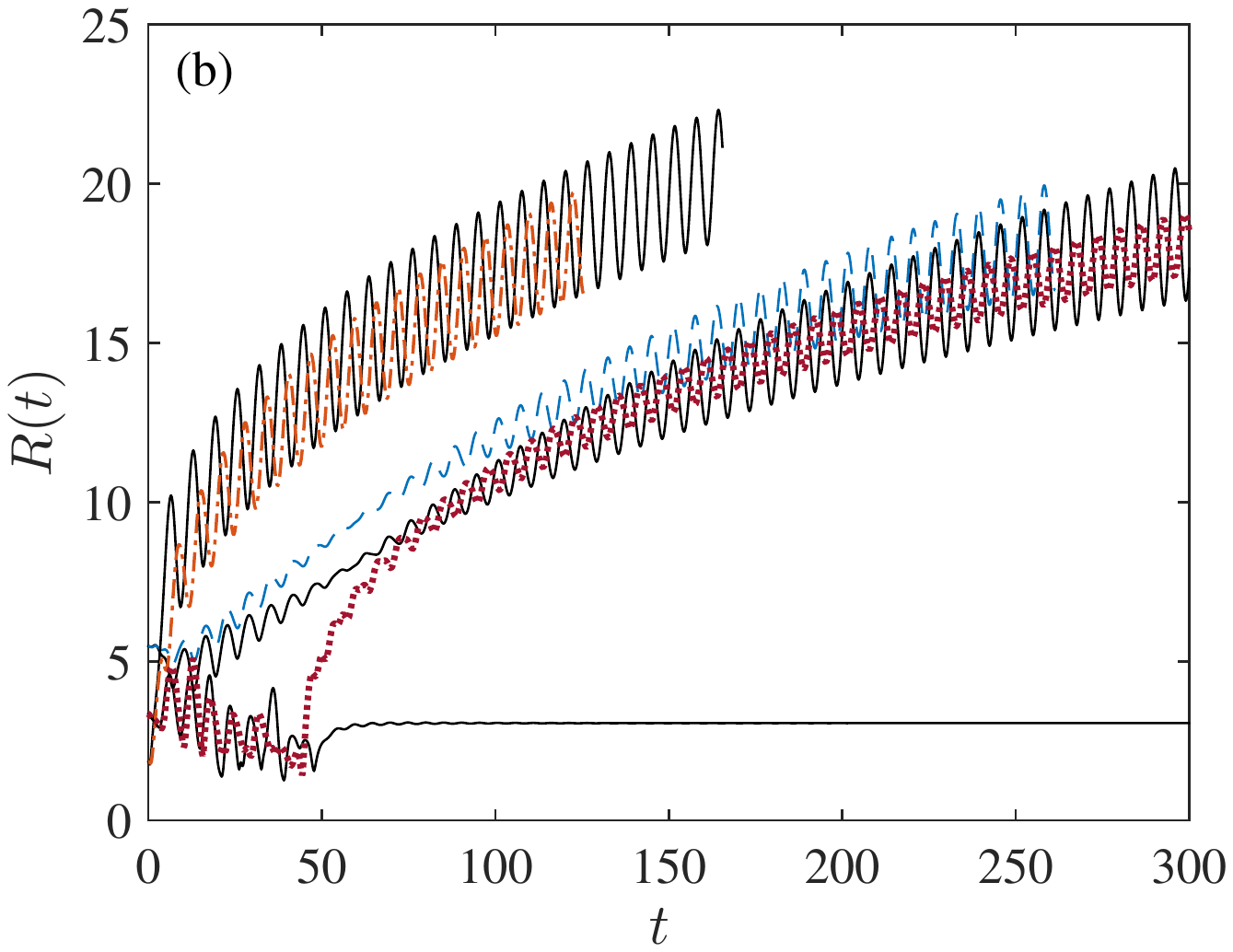}}}
\caption[]{Comparison between the evolution of pair separations
  obtained using the full pair-mobility matrix 
	(dashed, dotted and dash-dotted colored curves) and its
  multipole approximation (solid curves). Each panel
  presents three examples of pairs under time-dependent (a)
  and constant forcing (b). All pairs start from a fully aligned
  state. The multipole approximation includes the monopolar and
  dipolar terms.}
\label{fig:dipole}
\end{figure}

We present in Figs.\ \ref{fig:geometry:t}--\ref{fig:dipole} several examples
for the numerically integrated evolution of object pairs under various conditions.
One can be immediately appreciate the diversity of possible trajectories. 
To make your way through this richness it is important to make two distinctions between
types of trajectories. The first distinction is between 
constant forcing (as in sedimentation), which can make the objects only partially aligned
without synchronizing their phases of rotation~\cite{Gonzalez_etal2004,Krapf_etal2009},
and a time-dependent forcing protocol, which is known to lock the phase of an individual object onto
that of the force~\cite{Moths&Witten2013,Moths&Witten2013b}. 
The main issue examined below is how the presence of hydrodynamic interaction 
affects these two behaviors.
The second distinction, therefore, is whether hydrodynamic interactions are included 
(dashed, dotted and dash-dotted/colored curves) or turned off (solid gray curves).
In the absence of hydrodynamic interactions
(or when they get weak as the objects move far apart), the time-dependent aligning force
will make the objects fully synchronized, whereas under constant forcing 
the objects will generally become unaligned.

The results are presented in a dimensionless form, using units such that
$\eta=|\omega_0|=1$ and $\rho=0.01$. The distances between the stokeslets
of each object are taken randomly between $0$ and $1$; hence, $\rho \ll
l\sim 1$. The time-dependent forcing protocol is $\vec{F} = F_0
\left(-\sin(\omega_0t)\sin(\theta),\cos(\omega_0t)\sin(\theta),-\cos(\theta)
\right)$, where $\theta=0.1\pi$, $F_0=-|\lambda_3|^{-1}$,
$\omega_0=\text{sign}(\lambda_3)$ and $\lambda_3$ is the real eigenvalue of
the single-particle twist matrix.  We examine both the trajectories of
the separation vector connecting the origins of the two objects, and
the corresponding evolution of the orientation order parameter. 

We begin with the case of a time-dependent forcing,
Figs.\ \ref{fig:geometry:t} and \ref{fig:m:t}. The first observation,
most clearly demonstrated in Fig.\ \ref{fig:m:t}(b), is
that hydrodynamic interaction degrades the alignment of the two
objects, as has been rigorously inferred based on symmetry
considerations in Sec.\ \ref{sec:symmetry}. Another conclusion,
supported by additional examples not shown here, is that most objects,
which start sufficiently far apart, especially if they start fully
aligned, tend to repel each other (Fig.\ \ref{fig:geometry:t}). Even
if they are not fully aligned, the growing distance and weakening
interaction make them  individually  more aligned with the
forcing, and therefore also mutually synchronized.
Thus, the repulsion helps restore the alignment at long
times. The increasing separation occurs in the $xy$ plane, while along
the $z$ axis the separation decreases and saturates to a finite
distance, dependent on initial conditions, see Fig.\ \ref{fig:geometry:t}.

The repulsion is accompanied by a decrease in dissipation rate (up to small
oscillations), as demonstrated in Fig. \ref{fig:dissip}. When the HI
is turned off, the dissipation rate reaches a constant value as the
two independent objects set into their ultimate aligned state (solid
curves in Fig.\ \ref{fig:dissip}).

The repulsive effect is observed for most examples of our randomly generated
pairs of objects but is not a general law.
For instance, when the objects start at a sufficiently small separation,
some pairs remain ``bound'' in a limit cycle, oscillating about a
certain mean separation and mean orientational alignment, as
demonstrated by the green/dashed curves in Figs.\ \ref{fig:geometry:t} and \ref{fig:m:t}.

In Figs.\ \ref{fig:geometry:const} and \ref{fig:m:const} we examine
the same properties under constant forcing. The two effects--- degradation
in the alignment of a pair which is initially fully aligned, and
mutual repulsion--- are observed here as well. Yet, 
in the absence of a time-dependent aligning force, as the 
two objects move apart, alignment is not restored.
At long times, and for the common case of repulsion,
we distinguish between two observed behaviors:
a) The order parameter continues to change without saturating to a constant value 
(e.g., red/dash-dotted curve and cyan/dotted curve in Figs.~\ref{fig:m:const}(a) and \ref{fig:m:const}(b),
respectively).
This non-intuitive result can be explained as follows. The fact that the interaction becomes weak does not
necessarily imply that the accumulation of phase difference stops. If the two distant objects are partially aligned
we have $m\simeq \cos^2(\delta\omega_z t/2)$, where $\delta\omega_z$ is the difference between the objects' angular velocity
along the alignment axis. Hence, if the decay of $\delta\omega_z$ with time is slower then $t^{-1}$ then
phase difference will continue to accumulate. This depends on the detailed dynamics of repulsion which will be 
addressed in publication II.
b) The other option is that $m$ converges to some  value dependent on the initial state, with no particular chosen $m$ 
(green/dash-dotted curve in Fig.~\ref{fig:m:const}(b) and
cyan/dash-dotted curve in Fig.~\ref{fig:m:const}(c)),
i.e., the two objects continue to rotate with a fixed relative orientation.
In the examples that we checked there seems to be a tendency toward ultimate anti-alignment ($m=0$).
Therefore, we also checked the stability of anti-alignment 
in pairs which start from such a state. 
Fig.~\ref{fig:m:const}(d) examines the stability
of this configuration for objects initially confined to the $xy$ plane (perpendicular to the force).
Whereas the aligned pair (blue/dotted curve) is unstable, the anti-aligned one (red/dashed curve) remains stable
for the duration of integration.
It may well be that this stability survives for a long but finite time, see e.g., dark red/dotted curve
in Fig.~\ref{fig:m:const}(c). 
In addition, a separation of the pair along the $z$-axis destabilizes an anti-aligned pair as well (examples not shown). Finally, we note that even if the final phase difference were arbitrary and uniformly distributed, the value of $m$ would be evenly distributed around $1/2$ but non-uniformly, with larger weights on $m=0,1$. 
(This follows from the definition of $m$, see Eq.\eqref{eq:phi}.)

Figure \ref{fig:dipole} compares results obtained using the full
pair-mobility matrix of the stokeslet objects with those obtained from
the multipole (dipole) approximation. As expected, the two
calculations agree for objects whose mutual distance increases with
time, and disagree for objects whose trajectories reach close
proximity.

Further investigation (not shown here) of the orientational
dynamics of the objects suggests a possible explanation for the
characteristic repulsion between two chiral objects.  In the
absence of HI, each object rotates along $\hat{F}$ and translates
on average along $\hat{F}$. One contribution to the dipolar term of the HI
comes from the effect on each object by the vorticity of the Oseen flow 
caused by the other object.  This perturbative angular velocity is along an axis
which is perpendicular to the separation vector and the force,
$\hat{\omega}^a_{\rm flow} \propto -\hat{R} \times \hat{F}$ and
$\hat{\omega}^b_{\rm flow} \propto \hat{R} \times \hat{F}$. The
competition between this rotation and the aligning self-response of 
each individual object results in an inclination of the 
two objects relative to their non-interacting state. This
inclination alters the average unperturbed linear velocity of the object
by a small rotation about the $\hat{R} \times \hat{F}$ direction--- 
counter-clockwise for object $a$ and clockwise for object $b$. 
Hence, the two objects glide away from each other, 
$\dot{R}^2=2(\vec{V}^a-\vec{V}^b)\cdot\vec{R} \propto
((-\hat{R} \times \hat{F})\times\hat{F})\cdot\vec{R}
= R(1-(\hat{R}\cdot\hat{F})^2)\geq0$,
where the proportionality constant is positive, i.e., the separation increases with time
(unless $\vec{R}\parallel\vec{F}$, for which the whole argument does not hold).

\section{Discussion}
\label{sec:discuss}

Irregular objects display rich dynamics
already at the level of a pair of objects, as has been demonstrated
above. In the present work we have focused on the effect of the
hydrodynamic interaction on the orientational alignment of such object pairs.

The hydrodynamic interaction, in general, degrades the alignment. 
We have rigorously proven the instantaneous linear degradation for fully aligned objects at large mutual distances.
In other circumstances, such as nearby or unaligned objects, the hydrodynamic interaction
may have an opposite effect.
The leading degradation effect in distance is dipolar rather than monopolar; yet, it
is significant\,---\,a large mutual distance (compared to the object
size) is required to make the degradation negligible. More
quantitatively, the degradation will be significant when the
perturbation to the angular velocity due to HI, $\delta\omega$,
becomes comparable to the inverse of the time required to align a
single object. The unperturbed angular velocity is given by $\omega_0
= \Mmat{T}_{\rm self}F$. The dimensionless eigenvalue of the
self-twist matrix is generally found to be about an order of magnitude
smaller than the dimensionless self-mobility coefficient
\cite{Moths&Witten2013,Morozov&Leshansky2014,Moths&Witten2013b}, i.e.,
$\omega_0\sim 0.1 F/(8\pi
l^2)$. As presented in Sec.\ \ref{sec:symmetry}, $\delta\omega\sim
\Mmat{T}^{ab}F \sim F/(8\pi l^2) (l/R)^2$.
The alignment time is typically
$t_{\rm al} \sim 10/\omega_0$ (see Fig\ \ref{fig:m:t}). Hence, the
degree of degradation is $t_{\rm al}\delta\omega \sim 10^2(l/R)^2$. 
The conclusion is that the separation between the objects should be larger
than ten times their size to maintain alignment. In the case of many 
objects this implies a maximum volume fraction $(l/R)^3 \sim 10^{-3}$.

At the same time, as shown in Sec.\ \ref{sec:alignment}, for most of our randomly generated
pairs of objects,  
the hydrodynamic interaction makes the rotating objects repel each other. 
As a result, at long times the hydrodynamic interaction usually becomes negligible
and each of the objects gets aligned again with the time-dependent force.
In that section we also provided a possible explanation for this
repulsion, related to the mutual rotation of the two objects
which causes them to glide away from each other.
In fact, the objects need not be irregular to exhibit this gliding effect;
two forced ellipsoids which start parallel to one another
will experience the same repulsion~\cite{Claeys&Brady1993,Kutteh2010,Kim1986}. 
The resulting hydrodynamic ``pseudo-potential'' \cite{Squires&Brenner2000,Squires2001} 
will be addressed in a future publication. 
For the case of a finite number of objects
the repulsion will help restore the alignment as the objects drift
apart.
It should be kept in mind, however, that the repulsion is not a general law.
We observed it for a few dozens pairs of stokeslet objects. As mentioned above,
it also holds for a pair of well separated ellipsoids. Yet, a few counter-examples
have been also provided in Sec~\ref{sec:alignment}.

An interesting counterpart of the effects discussed here is found in the interaction between a forced 
object and a nearby wall \cite{Happel&Brenner,Russel_etal1977}.
The wall can be represented by an image (though not identical) object forced in 
the {\em opposite} direction~\cite{Pozrikidis}.  
As a result, the object will rotate and, if it is non-spherical, also glide {\em toward} the wall, as was indeed shown
for a rod falling near a wall~\cite{Russel_etal1977}.
Obviously, the interaction of an alignable object with a wall will also degrade the alignment.

An important distinction between regular and irregular objects,
which we have not dealt with here, concerns many-body interactions in forced systems.
A pair of forced spheres does not develop any
relative translational velocity \cite{Happel&Brenner}.
The same holds
for a pair of forced uniform ellipsoids to order $1/R^3$ (for an ellipsoid,
the components of $\Phi$ which correspond to the translational velocity vanish
\cite{BrennerIV}).
For a suspension of many objects this implies that two-body effects on
relative motion are either absent (spheres) or negligible at low
volume fraction (ellipsoids). By contrast, as we have shown here, a
pair of irregular objects develops a relative velocity
already at order $1/R^2$, which should lead to significant two-body interactions in
a suspension. This may bring about qualitative differences between driven
suspensions of regular and irregular objects in relation to such
phenomena as sedimentation.

This work shows that asymmetry in sedimenting objects leads to a wealth of hydrodynamic interaction effects not seen for spheres.  This study was undertaken to assess how interactions disrupt the rotational synchronization of such objects. However it proves to have striking effects independent of this alignment.  The prevalent repulsion, the occasional entrapment and the intricate quasiperiodic motions shown above are examples.  These effects could have significant impacts on real colloidal dispersions, e.g., in fluidized beds of catalyst particles. Though we have studied only pairwise interactions between identical objects, many of these effects are expected to apply more generally.   The general treatment of hydrodynamic interaction and its dependence on the
shape of the interacting objects, which we have developed here, should prove useful in exploring these phenomena.
Our work in progress aims to achieve a more general understanding of the rich behavior reported in Sec. \ref{sec:alignment}.

\begin{acknowledgments}
We thank Robert Deegan and Alex Leshansky for helpful discussions, and the James Franck Institute and 
Tel Aviv University for their hospitality during part of this work. 
This research has been supported by the US--Israel Binational Science Foundation
(Grant no.\ 2012090).
\end{acknowledgments}

\appendix
\numberwithin{equation}{section}

\section{Notation}
\label{sec:notation}

The dynamics of arbitrarily shaped objects is complex and requires an
elaborate notation. We use the following notation regarding vectors,
tensors, and matrices:
\begin{enumerate}

 \item 3-vectors are denoted by an arrow, $\vec{v}$, and unit 3-vectors
   by a hat, $\hat{v}$.

 \item 6-vectors are denoted by a calligraphic font, $\Cvvec{F}$.

 \item Matrices are marked by a blackboard-bold letter, e.g.,
   $\Mmat{M}$, where the dimension of the matrix is understood from
   the context.

 \item Tensors of rank 3 are denoted by a capital Greek letter, e.g., $\Phi$.
 
 \item A set of $N$ 3-vectors, representing $N$ stokeslets, is
   denoted by a bold letter, e.g., 
   $\bvec{v}^a=\left(\vec{v}^{a}_{1},\dots,\vec{v}^{a}_{N} \right)$.

 \item
  Subscripts with parentheses, e.g., $\Mmat{M}_{(2)}$, represent a term
  in a multipole expansion.

 \item
 $\Mmat{I}_{n\times n}$ is the $n\times n$ identity matrix.

 \item
Tensor multiplication\,---\,the dot notation, $\cdot$\,---\,denotes a
contraction over one index. The double dot notation, $:$, denotes a
contraction over two indices. Thus, given a tensor $\Upsilon$ of rank
$N$ and a tensor $\Xi$ of rank $M>N$, the tensors $\Upsilon \cdot \Xi$
and $\Upsilon : \Xi$ are tensors of rank $N+M-2$ and $N+M-4$.
For example, for $\Upsilon$ of rank 2 and $\Xi$ of rank 3,
$(\Xi \cdot \Upsilon)_{ikj}=\Upsilon_{is} \, \Xi_{skj}$ and
$(\Upsilon : \Xi)_{j}=\Upsilon_{ks} \, \Xi_{skj}$.

\item
The matrix $\vec{Y}^\times$ obtained from the vector $\vec{Y}$ is
defined as $(\vec{Y}^\times)_{ij}=\epsilon_{ikj}Y_k$, such that, for
any vector $\vec{X}$,
$\vec{Y}^\times\cdot\vec{X}=\vec{Y}\times\vec{X}$.

\end{enumerate}

\section{Pair-mobility: Change of object origin}
\label{sec:origin}

Here we derive the transformation of the pair-mobility matrix
under change of objects' origins.
Consider a new choice of origins given by $\vec{R}^{a \,\prime}=\vec{R}^{a} +\vec{h}^a$ and
$\vec{R}^{b \,\prime}=\vec{R}^{b} +\vec{h}^b$, and denote the objects' properties with respect to 
the new origins with~$^{\prime}$. 
Following Ref.~\citenum{Krapf_etal2009}, the transformations for the generalized velocities and forces
can be written as
$\Cvvec{V}^{x \,\prime}= [ \Mmat{I}_{6 \times 6}-(\Mmat{B}^x)^T ] \Cvvec{V}^{x}$
and  $\Cvvec{F}^{x \,\prime}= [ \Mmat{I}_{6 \times 6}+ \Mmat{B}^x ] \Cvvec{F}^{x}$ 
for $x=a,b$, where
$$
\Mmat{B}^a=
\begin{pmatrix}
 0	& 0 \\
  -\vec{h}^{a\times}  &0
 \end{pmatrix}
\text{ and }
\Mmat{B}^b=
\begin{pmatrix}
 0	& 0 \\
  -\vec{h}^{b\times}  &0
 \end{pmatrix} .
$$
Using $[ \Mmat{I}_{6 \times 6}+ \Mmat{B}^x ]^{-1}=[ \Mmat{I}_{6 \times 6}- \Mmat{B}^x ]$
we have
\begin{equation}
 \begin{pmatrix}
  \Mmat{M}^{\prime aa}	& \Mmat{M}^{\prime ab}\\
  \Mmat{M}^{\prime ba} & \Mmat{M}^{\prime bb}
 \end{pmatrix} =
 \begin{pmatrix}
 [\Mmat{I}_{6 \times 6}-(\Mmat{B}^a)^T]	& 0 \\
  0 &  [\Mmat{I}_{6 \times 6}-(\Mmat{B}^b)^T]
 \end{pmatrix}
 \begin{pmatrix}
  \Mmat{M}^{aa}	& \Mmat{M}^{ab}\\
  \Mmat{M}^{ba} & \Mmat{M}^{bb}
 \end{pmatrix}
\begin{pmatrix}
 [\Mmat{I}_{6 \times 6}-\Mmat{B}^a]	& 0 \\
  0 &  [\Mmat{I}_{6 \times 6}-\Mmat{B}^b]
 \end{pmatrix}.
\label{eq:origin_pair}
\end{equation}

\section{Properties of the tensor $\Phi$}
\label{sec:Phi}

Below we provide a more detailed discussion regarding the tensor $\Phi$
introduced in Sec.~\ref{sec:multipole}. We consider its symmetries
and its dependence on the choice of origin.
We separate $\Phi$ into 
a translational part--- linear velocity response to a flow gradient, denoted by
$\Phi_{\text{tran}}$, and
a rotational part--- angular velocity response to a flow gradient,
denoted by $\Phi_{\text{rot}}$. 
We show that $\Phi_{\text{tran}}$ is symmetric with respect to
its last two indices while 
$\Phi_{\text{rot}}$ has also an antisymmetric part which is the
Levi-Civita tensor. 
In addition, we show that $\Phi_{\text{tran}}$ depends
on the choice of the object's origin whereas $\Phi_{\text{rot}}$
does not, and derive the transformation of the former under change of origins.

In order to prove the symmetry properties of $\Phi$ we
consider its transpose tensor $\Phi^T=\tilde{\Phi}$ which gives the force dipole
around the object when subjected to external forcing,
$(\bvec{r}\bvec{F})=\tilde{\Phi}\cdot\Cvvec{F}=\tilde{\Phi}_{\text{tran}}\cdot\vec{F}
+\tilde{\Phi}_{\text{rot}}\cdot\vec{\tau}$.
We write the force dipole as a sum
of symmetric and anti-symmetric terms,
$\frac{1}{2}\left[(\bvec{r}\bvec{F})+(\bvec{r}\bvec{F})^T 
+ \epsilon\cdot\vec{\tau}\right]=\tilde{\Phi}_{\text{tran}}\cdot\vec{F}
+\tilde{\Phi}_{\text{rot}}\cdot\vec{\tau}$, where $\epsilon$
is the Levi-Civita tensor. The last equality implies 
that $(\tilde{\Phi}_{\text{tran}})_{ski}$ is symmetric with respect to $s$ and $k$
and that the anti-symmetric part of  $(\tilde{\Phi}_{\text{rot}})_{ski}$
is $\frac{1}{2}\epsilon_{ski}$.

Next we consider the transformation of $\Phi$ under change of origins.
Let us assume that an object is given in a constant, arbitrary shear flow
$\vec{u}(\vec{r})=S\cdot \vec{r}$, where $S$ is not necessarily a symmetric matrix. 
The object's linear velocities measured about $\vec{R}$ and 
$\vec{R}^{\prime}=\vec{R}+\vec{h}$ are 
$\vec{V}=S\cdot \vec{R}+\Phi_{\text{tran}} : S$ and 
$\vec{V}^{\prime}=S\cdot (\vec{R}+\vec{h})+\Phi_{\text{tran}}^{\prime} : S$ 
respectively.  
The tensor $\Phi_{\text{rot}}$ does not depend on the choice of origin since
the angular velocity of the object is independent of that choice,
$\vec{\omega}=\Phi_{\text{rot}} :S =\Phi^{\prime}_{\text{rot}} :S $.
Using the relation
$\vec{V}^{\prime }= \vec{V} -\vec{h} \times \vec{\omega}$ 
we find
$$
\Phi_{\text{tran}}^{\prime} : S =
(\Phi_{\text{tran}} - \vec{h}^{\times} \cdot \Phi_{\text{rot}} ) :S
-S\cdot \vec{h}.
$$
In general, with analogy to Eq.~\eqref{eq:origin_pair}, we can write
\begin{equation}
\Phi^{\prime}=[\Mmat{I}_{6 \times 6}-(\Mmat{B})^T]\cdot \Phi +\Delta,
\label{eq:Phi_origin}
\end{equation}
where
$$
\Mmat{B}=
\begin{pmatrix}
 0	& 0 \\
  -\vec{h}^{\times}  &0
 \end{pmatrix}
\text{ and }
\Delta_{iks}=  \left\{
  	\begin{array}{l l}
				-\delta_{is}h_k  \quad,i=1 \dots 3 \\
				0 \qquad,i=4 \dots 6						
		 \end{array}	    
	   \right..
$$

\section{Proofs of general properties of interaction multipoles}
\label{sec:proofs}

Here we prove the two general results presented in
Sec.~\ref{sec:multipole} concerning the interaction multipoles.

Multipole expansions are constructed by repeated projections
(``reflections''), between the two objects, of the Green's function
and its derivatives. The self-blocks of the mobility matrix result
from even projections, and the coupling blocks from odd
projections. In our case $\Mmat{G}$, the Oseen tensor, has even parity
and scales as $1/R$.

The Green's function $\Mmat{G}$ itself appears only once in the
expansion, in the first ($1/R$) multipole. This is because the force
monopoles acting on the particles are prescribed. This monopolar (odd)
interaction appears only in the coupling blocks. The leading multipole
appearing in the self-blocks is constructed by projecting the induced
force dipole on object 2 (proportional to $\nabla\Mmat{G}$) back onto
object 1 (by another $\nabla\Mmat{G}$). Thus, this leading multipole
is of 4th order, proportional to $1/R^4$. This proves the first result
in Sec.~\ref{sec:multipole}. Its particular manifestation for two
spheres is well known \cite{Happel&Brenner}.

Now, consider the $n$th multipole, proportional to $1/R^n$. Assume
that it contains $k$ $\Mmat{G}$'s and $n-k$ derivatives. Its parity is
$(-1)^{n-k}$. As explained above, for self-blocks $k$ is even, and
for coupling blocks it is odd. Hence, the parity of the $n$th
multipole is $(-1)^{n}$ in the self-blocks and $(-1)^{n+1}$ in the
coupling blocks. This proves the second result.

\section{General Form of $\Mmat{M}_{(2)}^{ab}$} 
\label{sec:general_form}

Below we provide a general form of the matrix $\Mmat{M}_{(2)}^{ab}$,
the 2nd-order multipole of the coupling block in the pair-mobility matrix, and
point out the number of its independent components.
This is done by decomposing the tensors $\Phi$ and $\Theta$ to their symmetric and
anti-symmetric parts.
Without loss of generality we choose the separation vector between the two objects
to be along the $x$ axis, $\hat{R}=\hat{x}$.
For two not necessarily identical objects the matrix $\Mmat{M}^{ab}_{(2)}$ has the form 
\begin{equation}
\Mmat{M}^{ab}_{(2)}=
\left( \frac{l}{R} \right)^2
\begin{pmatrix}
   \begin{pmatrix} 
     A_{xx}^a-A_{xx}^b & -A_{yx}^b & -A_{zx}^b\\
     A_{yx}^a &  0 & 0 	      \\
     A_{zx}^a & 0 & 0
   \end{pmatrix} &
   \begin{pmatrix} 
      -T_{xx}^b &  -T_{yx}^b & -T_{zx}^b\\
      0  & 0 & 1	      \\
      0 & -1 & 0
   \end{pmatrix}\\
   \begin{pmatrix} 
      T_{xx}^a &  0  & 0\\

      T_{yx}^a  & 0 & 1	      \\
      T_{zx}^a & -1 & 0
   \end{pmatrix}
    &0
\end{pmatrix},
\label{eq:general}
\end{equation}
where the $A^{x}_{ij}$ and $T^{x}_{ij}$ are functions of $\hat{R}$ and
the shape and orientation of object $x$, ($x=a,b$).
For two identical (in shape and orientation) objects
we have  
\begin{equation}
\Mmat{M}^{ab}_{(2)}=
\left( \frac{l}{R} \right)^2
\begin{pmatrix}
   \begin{pmatrix} 
    0 & -A_{yx} & -A_{zx} \\
     A_{yx} &  0 & 0 	      \\
     A_{zx} & 0 & 0
   \end{pmatrix} &
   \begin{pmatrix} 
      -T_{xx} &  -T_{yx} & -T_{zx}\\
      0  & 0 & 1	      \\
      0 & -1 & 0
   \end{pmatrix}\\
   \begin{pmatrix} 
      T_{xx} & 0  & 0\\
      T_{yx}  & 0 & 1	      \\
      T_{zx} & -1 & 0
   \end{pmatrix}
    &0
\end{pmatrix} .
\label{eq:general_id}
\end{equation}

\bibliography{pair_interaction_resubmit151105}

\end{document}